\documentclass[twocolumn]{aastex62}
\pdfoutput=1 
\usepackage{amsmath,amstext}
\usepackage[T1]{fontenc}
\usepackage[figure,figure*]{hypcap}
\usepackage{mwe}
\usepackage{lipsum}
\usepackage{graphicx} 
\usepackage{multirow}
\usepackage{xcolor,colortbl}

\newcommand{\twelveCO}{$^{12}$CO}
\newcommand{\thirtCO}{$^{13}$CO}

\newcommand{\eighteenCO}{C$^{18}$O}
\newcommand{\x}{$\times$}

\shorttitle{Physical Conditions of a Candidate Proto-Globular Cluster}
\shortauthors{Finn et al.}

\begin{document}

\title{New Insights into the Physical Conditions and Internal Structure of a Candidate Proto-Globular Cluster}

\author{Molly K. Finn}
\affiliation{Department of Astronomy, University of Virginia, Charlottesville, VA 22904, USA}

\author{Kelsey E. Johnson}
\affiliation{Department of Astronomy, University of Virginia, Charlottesville, VA 22904, USA}

\author{Crystal L. Brogan}
\affiliation{National Radio Astronomy Observatory, 520 Edgemont Road, Charlottesville, VA 22903, USA}

\author{Christine D. Wilson}
\affiliation{Department of Physics and Astronomy, McMaster University, 1280 Main St. W, Hamilton, ON L8S 4M1, Canada}

\author{Remy Indebetouw}
\affiliation{Department of Astronomy, University of Virginia, Charlottesville, VA 22904, USA}
\affiliation{National Radio Astronomy Observatory, 520 Edgemont Road, Charlottesville, VA 22903, USA}


\author{William E. Harris}
\affiliation{Department of Physics and Astronomy, McMaster University, 1280 Main St. W, Hamilton, ON L8S 4M1, Canada}

\author{Julia Kamenetzky}
\affiliation{Westminster College, 1840 S 1300 E. Salt Lake City, UT 84105 USA}

\author{Ashley Bemis}
\affiliation{Department of Physics and Astronomy, McMaster University, 1280 Main St. W, Hamilton, ON L8S 4M1, Canada}


\begin{abstract}

We present $\sim$ 0.1" resolution ($\sim$ 10 pc) ALMA observations of a molecular cloud identified in the merging Antennae galaxies with the potential to form a globular cluster, nicknamed the ``Firecracker.'' 
Since star formation has not yet begun at an appreciable level in this region, this cloud provides an example of what the birth environment of a globular cluster may have looked like before stars form and disrupt the natal physical conditions. 
Using emission from \twelveCO(2-1), \twelveCO(3-2), \thirtCO(2-1), HCN(4-3), and HCO$^+$(4-3) molecular lines, we are able to resolve the cloud's structure and find that it has a characteristic radius of 22 pc and a mass of 1--9$\times 10^6 M_\odot$.  
We also put constraints on the abundance ratios of \twelveCO/\thirtCO\ and H$_2$/\twelveCO. Based on the calculation of the mass, we determine that the commonly used CO-to-H$_2$ conversion factor in this region varies spatially, with average values in the range $X_{CO} = (0.12-1.1)\times10^{20}$ cm$^{-2}$ (K km s$^{-1}$)$^{-1}$. 
We demonstrate that if the cloud is bound (as is circumstantially suggested by its bright, compact morphology), an external pressure in excess of $P/k > 10^8$ K cm$^{-3}$ is required. 
This would be consistent with theoretical expectations that globular cluster formation requires high pressure environments, much higher than typical values found in the Milky Way. The position-velocity diagram of the cloud and its surrounding material suggests that this high pressure may be produced by ram pressure from the collision of filaments. 
The radial profile of the column density can be fit with both a Gaussian and a Bonnor-Ebert profile. If the Bonnor-Ebert fit is taken to be indicative of the cloud's physical structure, it would imply the cloud is gravitationally stable and pressure-confined.
The relative line strengths of HCN and HCO$^+$ in this region also suggest that these molecular lines can be used as a tracer for the evolutionary stage of a cluster.

\end{abstract}


\section{Introduction} \label{sec:intro} 

As some of the oldest objects in the universe, globular clusters are important probes of the early stages of galaxy formation and evolution. They are abundant in all massive galaxies \citep{Harris13}, despite theoretical predictions that they have a high mortality rate, with potentially $\lesssim 1\%$ surviving to 10 Gyr \citep{FallZhang01}. This suggests that the star formation process that created globular clusters was abundant in the early universe. 

The discovery of young, dense star clusters in nearby galaxies, dubbed ``super star clusters'' (SSCs), provided evidence that this star formation process is still occurring in the present universe \citep{OConnell94}. Further studies imply that these SSCs are likely very similar to the progenitors of the ancient globular clusters we are familiar with \citep{McLaughlinFall08}, though most will not survive to $>$10 Gyr. These clusters are primarily observed at optical and UV wavelengths, so most of our knowledge is confined to stages of evolution that occur after the progenitor cloud has formed stars and the cluster has at least partially emerged from its nascent molecular cloud. 

To observe the earliest stages of formation and evolution, we need to look at millimeter wavelengths that can see the structure of the molecular clouds, before stars have formed and while the birth environment is still intact. This stage of formation is expected to be short-lived, lasting only $\sim0.5-1$ Myr \citep{J15}, and so these objects are expected to be rare and therefore difficult to find. 

To form a globular cluster, a molecular cloud must have a sufficiently large mass within a relatively small radius. If we take the typical globular cluster to have a half-light radius of $\lesssim$10 pc \citep{vandenBergh91} and a stellar mass of $\gtrsim10^5 M_\odot$ \citep{HarrisPudritz94}, and assuming a star formation efficiency (SFE) of $20-50\%$ \citep{AshmanZepf01, Kroupa01}, then if a globular cluster loses approximately half its mass over the course of 10 Gyr, the progenitor molecular cloud must have an initial mass of $\gtrsim 10^6 M_\odot$ and a radius of $<25$ pc \citep{J15}. To constrain the evolutionary stage of the cluster to before the onset of star formation, the cloud must also have no associated thermal radio emission, which would penetrate the surrounding material and indicate that stars have formed and begun ionizing the surrounding gas.

We also expect that a molecular cloud forming a massive star cluster must be subject to a high external pressure. \cite{ElmegreenEfrefmov97} show that globular clusters with masses of $>10^5 M_\odot$ and core radii of 1-10 pc would require an external pressure of $P_0/k \sim 10^7 - 10^9$ K cm$^{-3}$ during formation for the resulting object to be bound. This pressure is orders of magnitude larger than typical ISM pressures in the disc of the Milky Way, and is likely to only be achieved in particular scenarios, including interactions between galaxy systems. This makes the merging Antennae galaxies, where high densities and pressures as well as an abundant population of optically-visible SSCs have been observed \citep{Whitmore00}, a prime location to search for such a molecular cloud. At a distance of 22 Mpc, it is also close enough that with ALMA, we are now able to resolve size scales that are comparable to those of the precursor molecular clouds which could generate globular clusters.

Using data from an ALMA Early Science project, \cite{Whitmore14} found a candidate pre-SSC cloud in the overlap region of the Antennae using CO(3-2) with a beam size of 0.56"$\times$0.43". Follow up analysis by \cite{J15} characterized it as having an inferred mass of 3.3--15$\times$10$^6 M_\odot$, a deconvolved radius of $<24\pm3$ pc, and a pressure of $P_0/k\gtrsim 10^9$ K cm$^{-1}$, all of which are consistent with expectations for a SSC-forming cloud. It also has no detectable associated thermal radio emission, where the upper limit on the peak ionizing flux from \cite{J15} is $N_\text{Lyc} \approx 6\times10^{50} \text{ s}^{-1}$, which corresponds to $\sim 60$ O-type stars, or ${\bf M_*} \lesssim 10^4 M_\odot$, which is more than two orders of magnitude less than the inferred mass of the cloud. Given that the expected resultant cluster will have a mass of $M_* >10^5 M_\odot$, this is taken to indicate that the Firecracker is likely to still be in a very early stage of formation. \cite{J15} also demonstrate that the cloud is most likely supported by turbulence, and so on a timescale of $\sim$ 1 Myr this turbulence will dissipate, initiating collapse if the cloud is bound, or dispersal if it is not. 

This cloud has been nicknamed the ``Firecracker,'' and to the best of our knowledge, is the only example found thus far that has the potential to be in the earliest stages of forming a massive star cluster with the potential to evolve into a globular cluster. Some very young SSCs have been identified with associated molecular gas \citep[e.g.][]{Leroy18,Turner17,Oey17}, but all of these also have associated thermal radio emission indicative of stars having formed. With the exception of one source from \cite{Leroy18}, this star formation is above the detection threshold for the Firecracker cloud.

\begin{figure*}
    \centering
    \includegraphics[width=\textwidth]{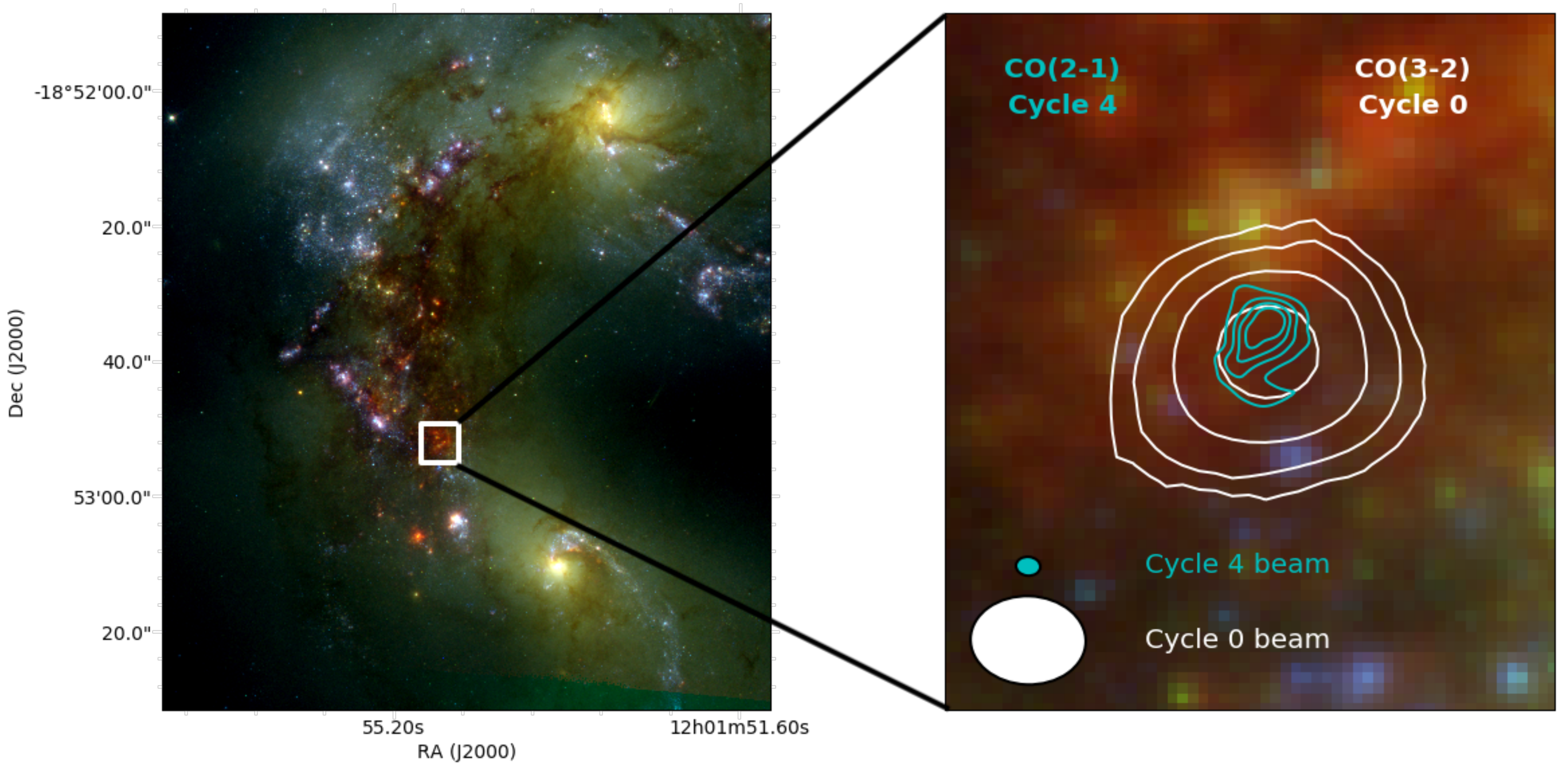}
    \caption{\emph{Left}: Three-color Hubble Space Telescope image of the Antennae galaxies where red is Pa$\alpha$, green is F814W, and blue is F435W. \emph{Right}: Proto-SSC in the Overlap region with CO(3-2) moment 0 contours (0.4, 0.8, 1.6, 3.2 Jy beam$^{-1}$ km s$^{-1}$, white) from \cite{J15} ALMA Cycle 0 data, and CO(2-1) moment~0 contours (10, 15, 20, 25$\sigma$, $\sigma = 0.04$ Jy beam$^{-1}$ km s$^{-1}$, cyan) from ALMA Cycle 4. The improved resolution of the Cycle 4 data allows us to now resolve the cloud and its structure. The synthesized beams for the Cycle 0 and Cycle 4 data are 0.43"$\times$0.56" (46$\times$60 pc) and 0.09"$\times$0.12" (10$\times$13 pc) respectively.}
    \label{fig:zoomin}
\end{figure*}

Here we present new, high resolution ALMA observations of \twelveCO(2-1), \twelveCO(3-2), and \thirtCO(2-1) emission that are capable of resolving the structure of the Firecracker cloud and improve upon the previous characterization of the source (Figure\,\ref{fig:zoomin}). The combination of the optically thick \twelveCO\ and the optically thin \thirtCO\ allow us to more directly measure the mass, while the improved resolution permits a more accurate size measurement for the cloud.

We also observe HCN(4-3) and HCO$^+$(4-3) emission, the ratio of which is postulated to be associated with evolutionary stage for massive cluster forming molecular clouds \citep{Johnson18}. These observations further confirm the Firecracker cloud is still in the early stages of evolution, with little disruption from star formation.

In Section\,\ref{sec:obs} we will discuss the ALMA observations that are used in this analysis. In Section\,\ref{subsec:lineprof}, we describe the extraction of the cloud from the surrounding medium, and in Sections \ref{subsec:mass}--\ref{subsec:Xco} we discuss obtaining the mass of the cloud and constraining the associated parameters. In Section\,\ref{subsec: NH2 profile}, we compare the column density structure to that predicted for a Bonnor-Ebert sphere. Section\,\ref{subsec:pressure} focuses on the pressure environment of the cloud and in Section\,\ref{subsec:PV} we consider cloud-cloud collision as a source of that pressure. In Section\,\ref{subsec:hcnhco} we discuss how HCN and HCO$^+$ can be used as tracers of the evolutionary state of cluster formation. In Section \,\ref{sec:discussion}, we discuss the various implications of these results for our understanding of cluster formation environments, and in Section\,\ref{sec:conclusions} we summarize the main findings in this work.

\section{Observations} \label{sec:obs}

We observed the overlap region of the Antennae galaxies using ALMA Band 6 and Band 7 in both extended and compact configurations during ALMA Cycles 3 and 4 (program codes 2015.1.00977.S and 2016.1.00924.S). The number of antennae online varied between 37 and 46. These observations are summarized in Table\,\ref{tab:obs}. The flux calibrators used were J1256-0547 and J1037-2934, and we estimate the flux uncertainty to be 10\% based on the variability in these sources. 
The bandpass was calibrated with J1256-0547, J1229+0203, and J1037-2934, and phase was calibrated with J1215-1731. These observations included continuum emission at each frequency, as well as emission from \twelveCO(2-1), \twelveCO(3-2), \thirtCO(2-1), \eighteenCO(2-1), HCN(4-3), HCO$^+$(4-3), CS(5-4), and H30$\alpha$.
The data from these observations were reduced and calibrated using the CASA 4.7.2 pipeline, and no self-calibration was performed. Images were created using Briggs weighting with robust parameters varying between 0.5 and 2.0, and using a 2646\x2646 pixel grid, with pixels of 0.014". The Firecracker cloud region is much smaller than the telescope primary beam (<1" compared to 16.9"-26.6" for our range of frequencies), so no primary beam correction is required.

\begin{table}
    \caption{ALMA Band 6 and Band 7 Observations of the Antennae}
    \begin{center}
    \begin{tabular}{c c c c}
        \hline
        \hline
        Date & Central & Time on  & Max.  \\
        & Freq. & Source & Baseline \\
        & (GHz) & (minutes) & (m) \\
        \hline
         Sep 17-22 2016 & 226 & 156 & 3200 \\
         Aug 8 2017 & 226 & 51 & 3700 \\
         Nov 22 2016 & 226 & 10 & 704 \\
         Aug 8 2017 & 237 & 30 & 3700 \\
         Nov 19 2016 & 237 & 9 & 704 \\
         Jul 23 2017 & 349 & 19 & 3600 \\
         Nov 27 2016 & 349 & 6 & 704 \\
         Dec 12 2016 & 349 & 5 & 650 \\
         \hline
    \end{tabular}
    \end{center}
    \label{tab:obs}
\end{table}

\begin{figure}
    \centering
    \includegraphics[width=0.48\textwidth]{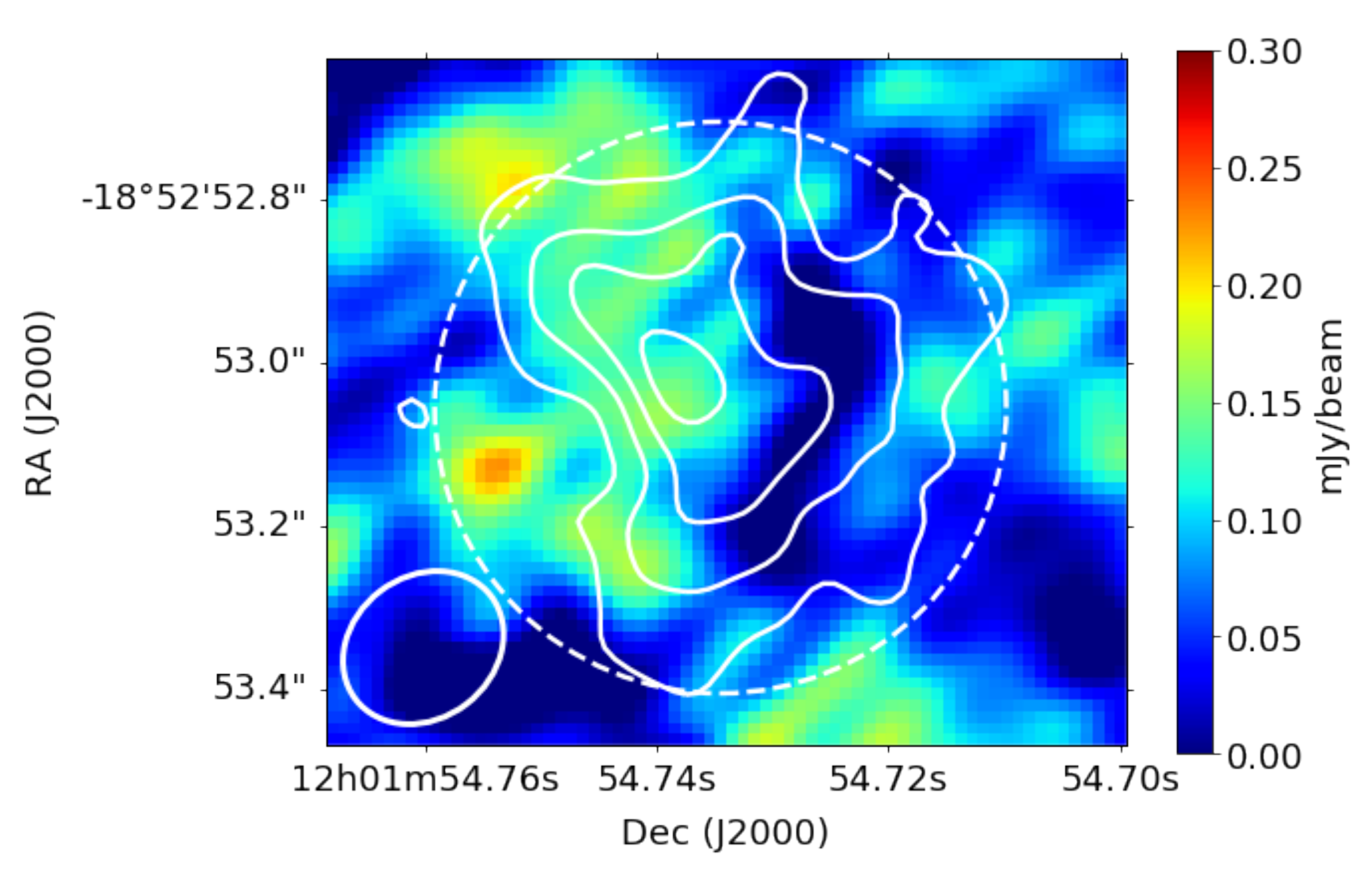}
    \caption{Band 7 continuum image, overlaid with contours of \thirtCO(2-1) and the 0.35" radius aperture in which the integrated flux was measured. The peak emission is 3.8$\sigma$ and is not coincident with the peak CO emission, leading to us consider this a non-detection. The image is scaled to 5$\sigma$, with $\sigma$ = 0.06 mJy beam$^{-1}$. The integrated flux within the white dashed aperture is $S_{880} = {0.78}\pm0.2$ mJy, which is likely due to the diffuse continuum emission in the region, and is used to set an upper limit on the mass of the cloud. The synthesized beam is shown in the lower left corner and has a size of 0.17"\x0.21".}
    \label{fig:B7cont}
\end{figure}

In the vicinity of the Firecracker, there is diffuse continuum emission at all three frequencies throughout the area, associated with the SGMCs in the overlap region. However, there is no peak in emission or any morphology in the continuum associated with the Firecracker itself, based on the well-detected CO emission (see Figure\,\ref{fig:B7cont}). We therefore consider this a non-detection of the Firecracker, with 5$\sigma$ upper limits for the peak emission of $3.0\times10^{-4}$ Jy beam$^{-1}$ at 349 GHz, $9.5\times10^{-5}$ Jy beam$^{-1}$ at 237 GHz, and $6.0\times10^{-5}$ Jy beam$^{-1}$ at 226 GHz.
We also did not detect \eighteenCO(2-1), CS(5-4), or H30$\alpha$ in this region. The $5\sigma$ upper limits for these transitions are 0.85 mJy beam$^{-1}$ for \eighteenCO(2-1), 1.75 mJy beam$^{-1}$ for CS(5-4), and 2.25 mJy beam$^{-1}$ for H30$\alpha$ at a velocity resolution of 10 km s$^{-1}$ for each.

\begin{figure}
    \centering
    \includegraphics[width=0.47\textwidth]{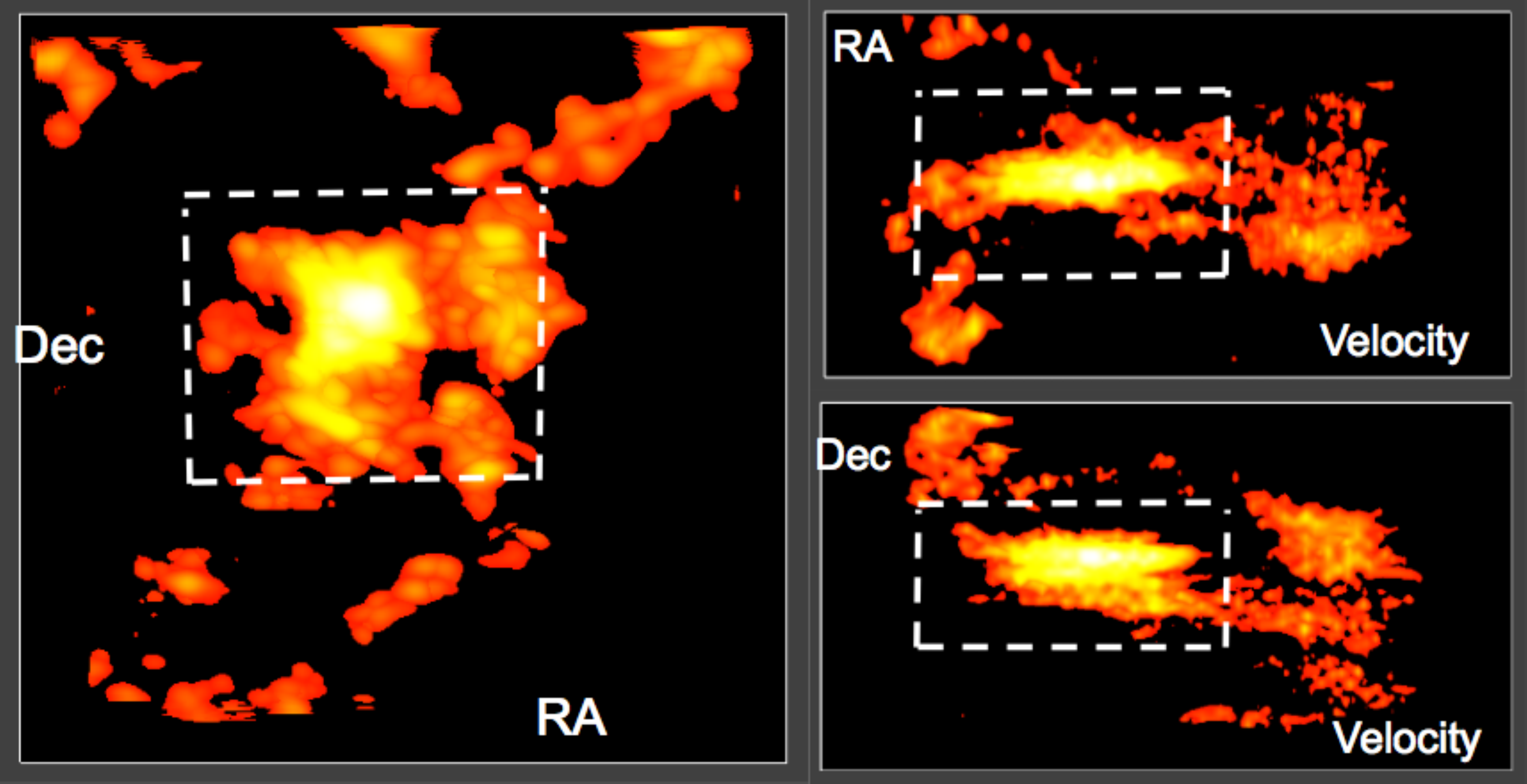}
    \caption{Three different 2-D projections of the \twelveCO(2-1) data cube using the 3-D visualization tool \texttt{shwirl} to examine the structure of the cloud in velocity space. Each box shows the same range of data. The white dashed line shows the extraction of the cloud from the surrounding field and second velocity component. The extraction is a cube extending 0.98" in right ascension, 0.84" in declination, centered on 12:01:54.73~-18:52:53.1 and with a velocity range of 1430--1555 km s$^{-1}$. }
    \label{fig:shwirl}
\end{figure}

For the remaining transitions, the Firecracker cloud was detected strongly in \twelveCO(2-1), \twelveCO(3-2), and \thirtCO(2-1), and weakly in HCN(4-3) and HCO$^+$(4-3). The parameters of the data cubes for each detected transition are summarized in Table \ref{tab:datacubes}. 
The RMS was determined using line-free channels. Figure\,\ref{fig:lineprof} shows the three CO line profiles in this region, and Figure \ref{fig:shwirl} shows the full \twelveCO(2-1) emission cube's spatial and velocity structure.

\begin{table}
    \begin{center}
    \caption{Data cube parameters for detected transitions in the Firecracker region}
    \begin{tabular}{c c c c c}
        \hline
        \hline
         Transition & Robust & Synth. Beam & RMS/chan & Channel \\
          & & (arcsec$^2$) & (mJy/beam) & (km/s) \\
         \hline
         \twelveCO(2-1) & 0.5 & 0.09\x0.12 & 0.6 & 5 \\
         \twelveCO(3-2) & 0.5 & 0.15\x0.16 & 2.0 & 5 \\
         \thirtCO(2-1) & 2.0 & 0.17\x0.18 & 0.25 & 5 \\
         HCN(4-3) & 2.0 & 0.17\x0.20 & 1.2 & 15 \\
         HCO$^+$(4-3) & 2.0 & 0.17\x0.21 & 1.0 & 15 \\
         \hline
    \end{tabular}
    \end{center}
    \textbf{Notes.} Quoted channel widths reflect values adopted to improve sensitivity.
    \label{tab:datacubes}
\end{table}

\section{Analysis} \label{sec:analysis}

\subsection{CO Line Profiles and Cloud Extraction \label{subsec:lineprof}}

\begin{figure}
    \centering
    \includegraphics[width=0.45\textwidth]{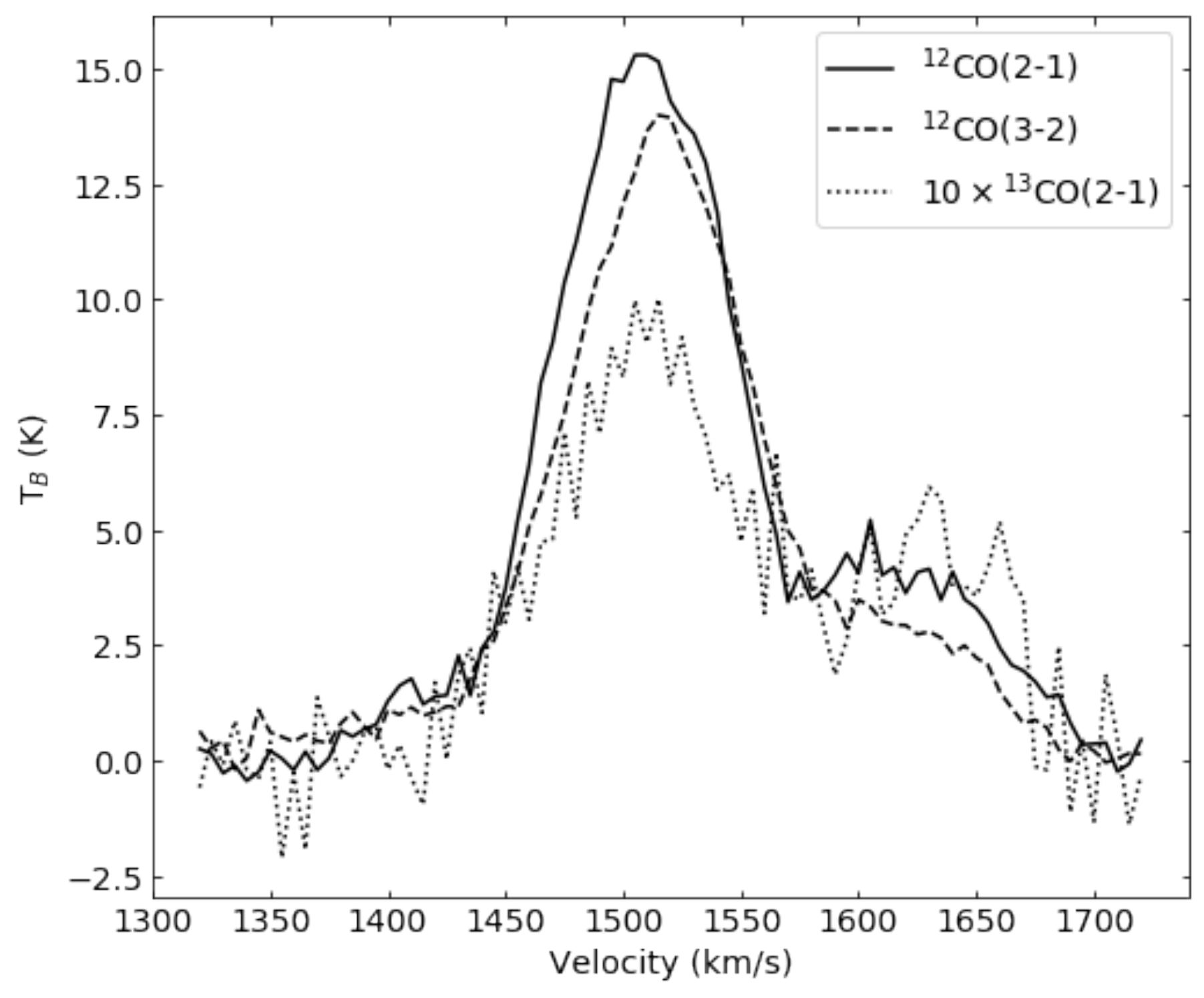}
    \caption{Line profiles of \twelveCO(2-1), \twelveCO(3-2), and \thirtCO(2-1) as measured in a 0.24" radius region centered on the peak \thirtCO(2-1) emission. The \thirtCO(2-1) profile has been multiplied by a factor of 10 for easier comparison. Each data set was convolved to the same beam size of 0.17"$\times$0.18". The similarity of the peak brightness temperatures of \twelveCO(2-1) and \twelveCO(3-2) indicate that these lines are approximately thermalized. Furthermore, these spectra show that there is a second velocity component along the line of sight, as expected from previous observations by \cite{J15}. This component is assumed to be separate from the Firecracker cloud, so we extract only the primary velocity component from the data cube for all further analysis (see Figure\,\ref{fig:shwirl}).}
    \label{fig:lineprof}
\end{figure}

 To obtain and compare line profiles for the \twelveCO(2-1), \twelveCO(3-2), and \thirtCO(2-1) emission, the data cubes were each convolved to the synthesized beam of \thirtCO(2-1) and are measured within a 0.24" radius aperture. The peak brightness temperatures of \twelveCO(2-1) and \twelveCO(3-2) are $T_{CO(2-1)} = 17\pm2$ K and $T_{CO(3-2)} = 16\pm2$ K, suggesting that these transition lines are nearly thermalized in this region. These line profiles are shown in Figure \ref{fig:lineprof}.

 From these profiles, we also see that there is a second velocity component along the line of sight that we infer is a separate cloud that should not be included in analysis. Using the 3D visualization tool \texttt{shwirl} \citep{shwirl}, we show the extraction of the cloud from the surrounding field. This corresponds to a 0.98"\x0.84" rectangle around the cloud centered on 12:01:54.73~-18:52:53.1, and velocities in the range 1430--1555 km s$^{-1}$.

\begin{figure*}[t]
    \centering
    \includegraphics[width=\textwidth]{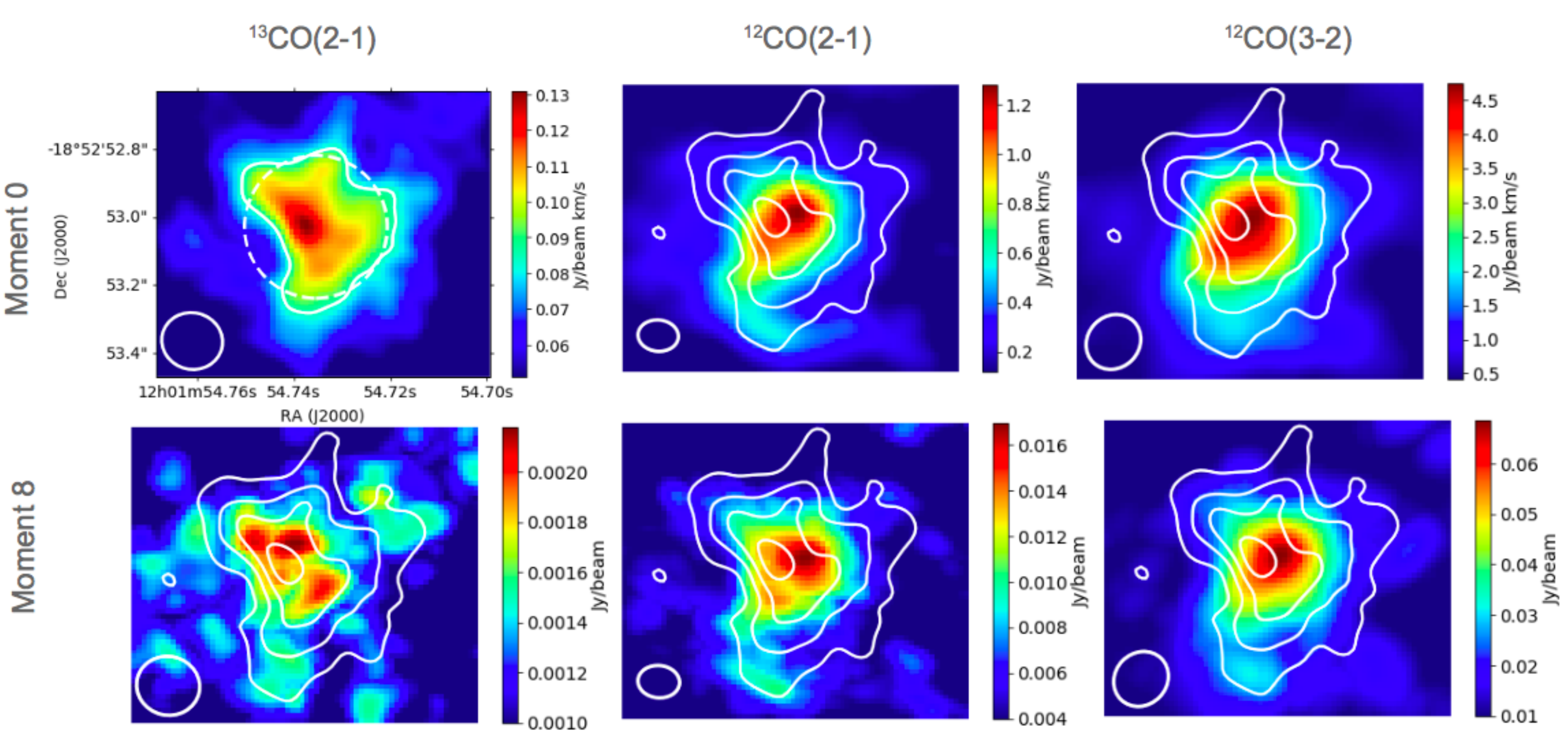}
    \caption{\emph{Top row}: Total intensity (moment~0) maps. \emph{Bottom row}: Peak intensity (moment~8) maps. \emph{Left column}: \thirtCO(2-1). \emph{Middle column}: \twelveCO(2-1). \emph{Right column}: \twelveCO(3-2). In the \thirtCO(2-1) moment~0 image (upper left), the solid white line represents the 5$\sigma$ ($\sigma$=0.017 Jy beam$^{-1}$ km s$^{-1}$) contour, and the dashed white line represents the circle with area equal to that enclosed by the 5$\sigma$ contour. The radius of this circle is 0.21", which is taken to be the characteristic radius of the cloud. In the other images, contours of the \thirtCO(2-1) are overplotted at 4, 5, 6, and 7$\sigma$ levels. Synthesized beams are shown in the bottom left corners of each image, and are 0.17"$\times$0.18" for \thirtCO(2-1), 0.09"$\times$0.12" for \twelveCO(2-1), and 0.15"$\times$0.16" for \twelveCO(3-2).}
    \label{fig:moments}
\end{figure*}

From this extraction, we made total intensity maps (moment~0), integrating over the velocity range 1430-1555 km s$^{-1}$, and peak intensity maps (moment~8) for \twelveCO(2-1), \twelveCO(3-2), and \thirtCO(2-1). 
The total intensity (moment~0) and peak intensity (moment~8) maps for each CO transition are shown in Figure \ref{fig:moments}. The properties of the cloud measured from these cubes for the extracted region are given in Table \ref{tab:cloud_params_measured}.

To determine the size of the cloud from these observations, we use the \thirtCO(2-1) emission, since it is optically thin and gives a better representation of the cloud's structure than the optically thick \twelveCO.
We define a characteristic radius, which is the radius of a circle with the same area as that enclosed by the 5$\sigma$ contour of the \thirtCO(2-1) total intensity (moment~0) map. This characteristic radius for the Firecracker is 0.21", which corresponds to a size of 21 pc at a distance of 22 Mpc.

\begin{table*}
    \begin{center}
    \caption{Molecular Cloud Measured Properties}
    \begin{tabular}{c c c c c c c c c}
        \hline 
        \hline 
        RA & Dec & V$_{LSR}$ & S$_{^{12}CO(2-1)}$ & S$_{^{12}CO(3-2)}$ & S$_{^{13}CO(2-1)}$ & $\sigma_{V,^{12}CO(2-1)}$ & $\sigma_{V,^{12}CO(3-2)}$ & Radius \\
        (J2000) & (J2000) & (km s$^{-1}$) & (Jy km s$^{-1}$) & (Jy km s$^{-1}$) & (Jy km s$^{-1}$) & (km s$^{-1}$) & (km s$^{-1}$) & (arcsec) \\
        \hline 
        12:01:54.73 & -18:52:53.0 & 1516$\pm$5 & 14$\pm$1 & 26$\pm$2 & 0.87$\pm$0.2 & 36$\pm$3 & 38$\pm$3 & 0.21 \\ 
        \hline
    \end{tabular}
    \end{center}
    \textbf{Notes.} Measurements of the velocity properties were based on Gaussian fitting of the line profiles for \twelveCO(2-1) and \twelveCO(3-2), after the second velocity component was subtracted out with a Gaussian fit. The integrated flux was measured within an aperture with a radius of 0.35". The characteristic radius is the radius of a circle with the same area as that enclosed by the 5$\sigma$ contour of the \thirtCO(2-1) moment~0 map (see Figure\,\ref{fig:moments}). 
    \label{tab:cloud_params_measured}
\end{table*}


\subsection{Cloud Mass \label{subsec:mass}}


Observations of both \twelveCO(2-1) and \thirtCO(2-1) allow us to determine the optical depth of the cloud by assuming an abundance ratio for these two molecules and assuming that their excitation temperatures are the same. If the excitation temperature of \thirtCO\ is lower than \twelveCO, this assumption will underestimate the mass. We also assume that \twelveCO(2-1) is thermalized with respect to \twelveCO(1-0), as expected from the line profiles (Figure\,\ref{fig:lineprof}). If \twelveCO(2-1) is not thermalized, this assumption will also underestimate the mass. 

In the overlap region of the Antennae, the \twelveCO/\thirtCO\  abundance ratio has been measured to be $X_{12} / X_{13} \simeq 70$, though it is poorly constrained in this region and could vary from 40 to 200 \citep{Zhu03}. We convolved the \twelveCO(2-1) image to the synthesized beam of \thirtCO(2-1), then fit Gaussian profiles to the velocity profile for each \twelveCO(2-1) pixel. We then fit Gaussian profiles to the \thirtCO(2-1) velocities, fixing the central velocity to be the same as the corresponding \twelveCO(2-1) pixel, and masking pixels where a solution to the fit could not be found. Taking the ratio of the peak brightness temperatures of these two molecular lines at each unmasked pixel, we created a map of the peak optical depth, $\tau_{12}$, using the equation

\begin{equation}
    \frac{T_{12}}{T_{13}} = \frac{T_{x,12}}{T_{x,13}}\frac{1 - e^{-\tau_{12}}}{1 - e^{-\tau_{13}}} = \frac{1 - e^{-\tau_{12}}}{1 - e^{-\tau_{13}}},
    \label{eq:tau12}
\end{equation}

\noindent by taking $\tau_{12} / \tau_{13} = X_{12} / X_{13}$ and assuming that the excitation temperatures for \twelveCO\ and \thirtCO\ are equal ($T_{x,12} = T_{x,13}$). From these optical depths, we found the peak excitation temperature, $T_x$, at each pixel given by

\begin{equation}
    T_{12} = (1 - e^{-\tau_{12}})\frac{T_{\text{UL}}}{e^{T_{\text{UL}}/T_x} -1 },
    \label{eq:Tx}
\end{equation}

\noindent where $T_{\text{UL}}=11.07$ K for \twelveCO. 

We then determined $\tau_{12}$ for each velocity in the cube, giving us a profile of the optical depth for each pixel. This was done by using Equation\,\ref{eq:tau12} with the ratio of the brightness temperature of \twelveCO(2-1) and \thirtCO(2-1) wherever \thirtCO(2-1) was detected ($>4\sigma$, $\sigma$ = 0.02 K). Where \thirtCO(2-1) was not detected and \twelveCO(2-1) was detected, we used Equation\,\ref{eq:Tx} with the brightness temperature of \twelveCO(2-1) only, and assuming that the excitation temperature does not vary with velocity. The data cubes were masked with thresholds of $3\sigma$ ($\sigma$ = 0.6 mJy beam$^{-1}$ for \twelveCO(2-1) and $\sigma$ = 0.25 mJy beam$^{-1}$ for \thirtCO(2-1)). With the assumption that the excitation temperature remains constant at all velocities within the cloud, we found the column density at each pixel using the equation from \cite{MangumShirley15}:

\begin{equation}
    N_{tot} = \frac{8\pi \nu_0^2 Q}{c^2 A_{ul} g_u} e^{\frac{E_u}{kT_{x}}}\left(e^\frac{h\nu_0}{kT_{x}} - 1 \right)^{-1} \int_0^\infty \tau_\nu d\nu
\end{equation}

For \twelveCO, this equation becomes

\begin{equation}
    \frac{N^{12}_{tot}}{\text{cm}^{-2}} = 3.3\times 10^{14} \left(\frac{T_{x}}{B_0} + \frac{1}{3}\right) \frac{1}{e^{\frac{-5.53}{T_{x}}} - e^{\frac{-16.6}{T_{x}}}} \int_0^\infty \tau_v dv,
    \label{eq:Ntot}
\end{equation}

\noindent where $B_0 = 2.7674$ K for \twelveCO. 

\begin{figure}
    \centering
    \includegraphics[width=0.47\textwidth]{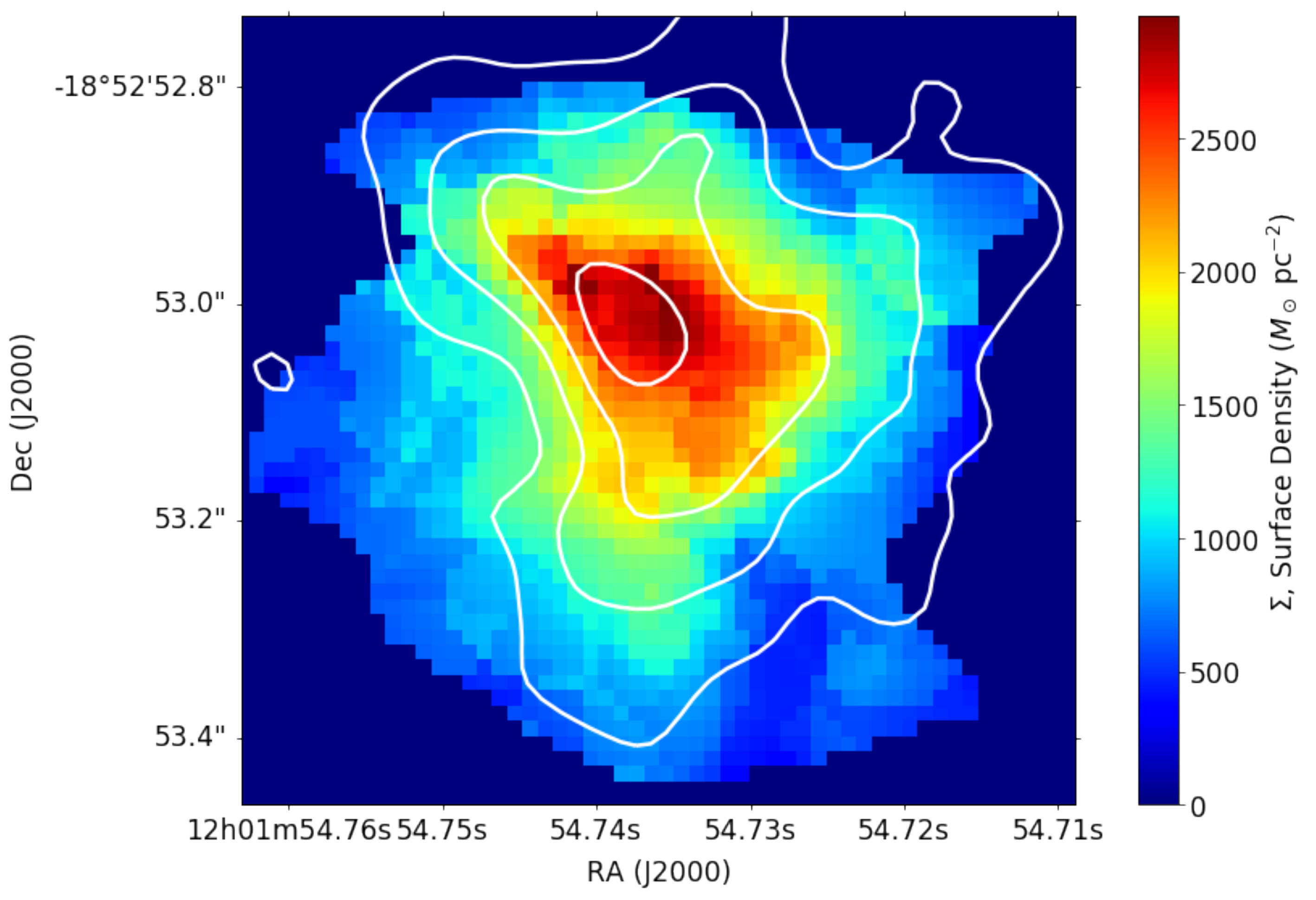}
    \caption{Map of the mass surface density. This version was created with the assumed parameters $X_{12} / X_{13}=70$ and H$_2$/\twelveCO$=10^{4.5}$. Summing over all the pixels and assuming a pixel area of 2.23 pc$^2$ results in a total mass of $4.5\times10^6 M_\odot$. Overplotted are contours of \thirtCO(2-1) moment 0.}
    \label{fig:massmap}
\end{figure}

From these column densities, we take an H$_2$/\twelveCO\ abundance ratio of H$_2$/\twelveCO\ = $10^4-10^5$, which is typical of Milky Way IRDCs, before or slightly after protostellar objects have formed, akin to the stage we expect the Firecracker to most likely be in \citep{Gerner14}. The Antennae has a very nearly solar metallicity, [Z] = +0.07 $\pm$ 0.03 \citep{Lardo15}, so the default expectation is that its chemistry is similar to the Milky Way's. We then assume the total mass is 1.3 times the mass of H$_2$ to derive the total mass surface density (shown as a map in Figure\,\ref{fig:massmap}). Taking a pixel area of $A$ = 2.23 pc$^2$, we can add the mass from each pixel to get the total mass of the cloud. 

Different combinations of values in the expected ranges of $X_{12} / X_{13}$ and H$_2$/{\twelveCO} were used, resulting in masses that varied in the range 1.0--31\x10$^6 M_\odot$. In Section\,\ref{subsec:contemission}, we will put additional constraints on the upper limit of this mass range due to dust emission. Mass estimates for a few selected parameter combinations are given in Table \ref{tab:masses}, and estimates for the full range of parameter combinations are shown in Figure\,\ref{fig:abunparams}. The mass directly tracks variations in H$_2$/{\twelveCO}, with an order of magnitude change in H$_2$/{\twelveCO} corresponding to an order of magnitude change in mass. The resulting mass is less sensitive to $X_{12} / X_{13}$, with a factor of five change in this value only resulting in a factor of $\sim3$ change in mass. 

The range in masses that results from varying these parameters is in good agreement with measurements made by \cite{J15}, which had lower spatial resolution (0.56"$\times$0.43") and did not have optical depth information.

\definecolor{Gray}{gray}{0.92}
\newcolumntype{g}{>{\columncolor{Gray}}c}

\begin{table}
    \begin{center}
    \caption{Possible values for the mass ($10^6 M_\odot$)}
    \begin{tabular}{c c | c c c c}
        \hline
        \rowcolor{white}&&\multicolumn{4}{c}{$X_{12} / X_{13}=$}  \\
         &&  40 & 70  &  120 &  200 \\
         \hline
         \hline
         \multirow{3}{*}{H$_2$/\twelveCO\ = }
         & $10^4$ & 1.0 & 1.4 & 2.1 & 3.1 \\
         & $10^{4.5}$ & 3.3 & 4.5 & 6.5 & 9.7 \\
         & $10^5$ & 11 & 14 & 21 & 31 \\
         \hline
    \end{tabular}
    \end{center}
    \textbf{Notes.} Masses for given combinations of $X_{12} / X_{13}$ and H$_2$/\twelveCO\ assumptions are given in the body of the table with units of $10^6 M_\odot$. 
    \label{tab:masses}
\end{table}

\subsection{Expected Continuum Emission \label{subsec:contemission}}

The lack of detected continuum emission associated with the Firecracker sets further constraints on the mass of the cloud. At all three frequencies, there is diffuse continuum emission associated with the larger region, but no morphology or peak emission associated with Firecracker above $3.8\sigma$ (Figure\,\ref{fig:B7cont}). We consider this a non-detection of the Firecracker, and use the integrated flux from the diffuse emission to set an upper limit on the Firecracker's dust mass. 

Using the Band 7 observations (in which the dust emission from the Firecracker should be brightest), we flagged the emission lines to create a continuum image with a beam FWHM of 0.17"\x0.21" and an RMS of 0.06 mJy beam$^{-1}$. The integrated flux in a 0.35" radius circular region around the Firecracker is $S_{880} = {0.78}\pm0.2$ mJy. 

From \cite{Wilson08},

\begin{equation}
    M_{dust} = 74,220 S_{880} D^2 \frac{(e^{17/T} - 1)}{\kappa} (M_\odot),
\end{equation}

\noindent where $S_{880}$ is measured in Janskys at 880$\mu$m (Band 7), $D$ is measured in Mpc, $\kappa$ is the dust emissivity measured in cm$^2$ g$^{-1}$, and $T$ is measured in K. For the Antennae system, $D = 22$ Mpc. Taking $T_{Kin}\simeq T_{ex}$, the temperature measured in this region is 25--35 K. Typical values adopted for the dust emissivity and gas-to-dust ratio in these types of environments are $\kappa=0.9\pm0.13$ cm$^2$ g$^{-1}$ and a ratio of $120\pm28$ \citep{Wilson08}. 

If we take the most extreme values to maximize $M_{dust}$ within the expected range for each parameter (so $S_{880} \leq 1.34$ mJy, $T \geq 25$ K, $\kappa \geq 0.77$ cm$^2$ g$^{-1}$), the upper limit on the dust mass would be $M_{dust} \leq 6\times10^4 M_\odot$. Taking the maximum gas-to-dust ratio $ \leq 148$, the largest total mass that would be consistent with the continuum non-detection would be $9\times10^6 M_\odot$. This is then taken as the upper limit on the mass of the Firecracker.

We also note that this continuum non-detection at 880$\mu$m would suggest that the previous unresolved continuum detection at this frequency in \cite{J15} was likely instead picking up the diffuse emission of the larger region rather than the Firecracker itself.

\begin{figure}
    \centering
    \includegraphics[width=0.45\textwidth]{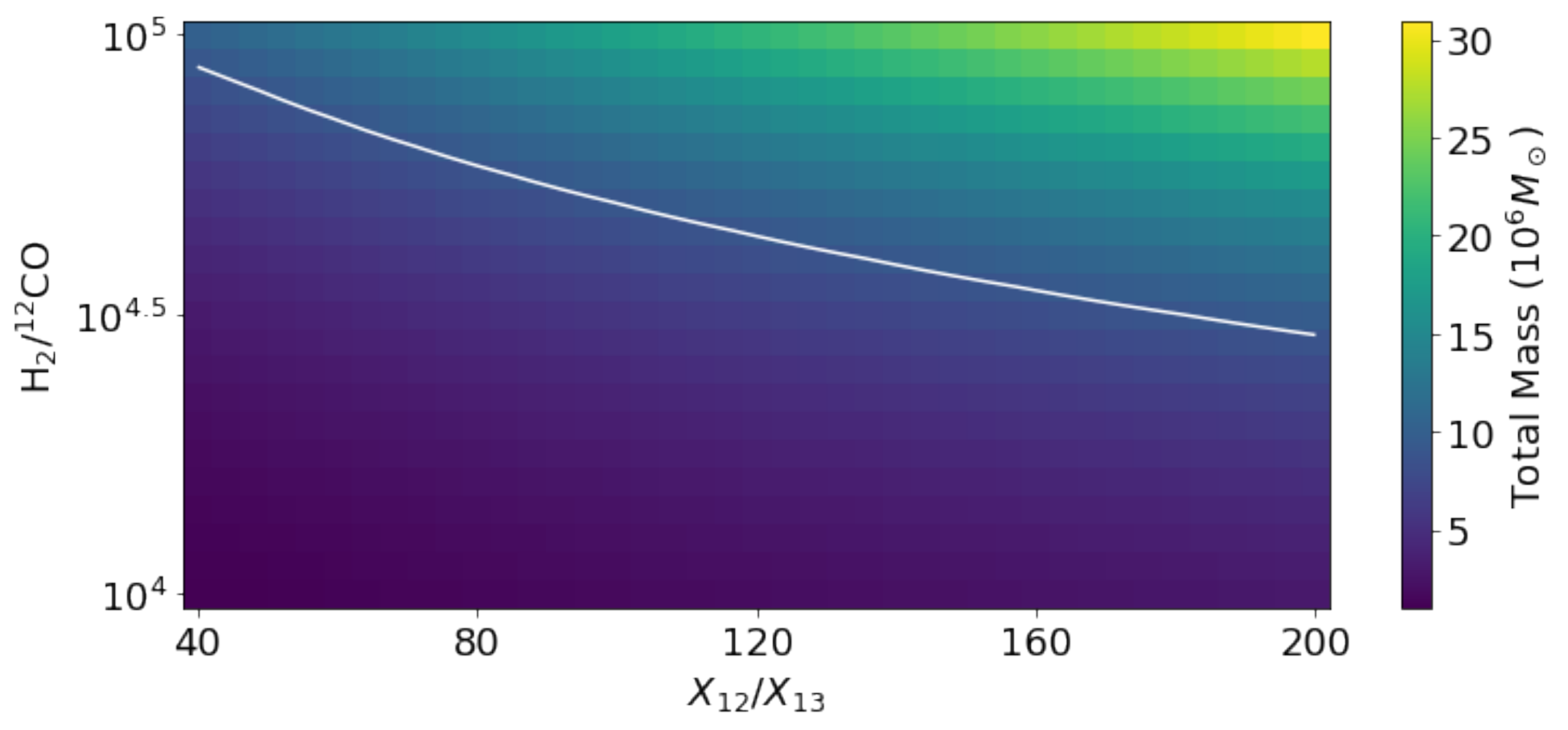}
    \caption{Mass estimates derived from the full ranges of expected $X_{12} / X_{13}$ and H$_2$/\twelveCO\ values. The white line represents the upper limit on the mass derived from the continuum non-detections ($9\times10^6\ M_\odot$), and so parameter combinations falling above this line on the plot can be ruled out for the Firecracker region. The lower limit on the mass ($1\times10^6\ M_\odot$) is set by the lower limits of the adopted $X_{12} / X_{13}$ and H$_2$/\twelveCO\ values in the bottom left of the plot.} 
    \label{fig:abunparams}
\end{figure}

We can compare this limit to the mass estimates from different combinations of  $X_{12} / X_{13}$ and H$_2$/{\twelveCO} values. Mass estimates for the full range of parameter combinations are shown in Figure\,\ref{fig:abunparams}, with a line representing the upper limit derived from the lack of continuum emission. Parameter combinations above this line are ruled out for the Firecracker region.  For example, if the assumed $X_{12} / X_{13}$ ratio is taken to be 200, the value of H$_2$/\twelveCO\ must be less than $10^{4.5}$.

\subsection{$X_{CO}$ Conversion Factor \label{subsec:Xco}}

We use the masses and column densities that we derive to calculate the CO-to-H$_2$ conversion factor, $X_{CO}$, in the Firecracker. In starburst regions, this conversion factor is typically taken to be $X_{CO} = 0.5\times10^{20}$ cm$^{-2}$ (K km s$^{-1}$)$^{-1}$, but is expected to vary by up to a factor of four \citep{Bolatto13}. In non-starbusting regions, the typical value taken is $X_{CO} = 2.0\times10^{20}$ cm$^{-2}$ (K km s$^{-1}$)$^{-1}$.

Considering the range of expected masses up to the upper mass limit from the continuum non-detections, we create maps of $X_{CO}$ in the Firecracker cloud for each set of abundance parameter assumptions. Fitting a Gaussian to the distribution of values within each map, the average values vary in the range $X_{CO} = (0.12-1.1)\times10^{20}$ cm$^{-2}$ (K km s$^{-1}$)$^{-1}$ for the Firecracker region. This is consistent with the typically assumed value for starbursts.

\begin{figure}
    \centering
    \includegraphics[width=0.45\textwidth]{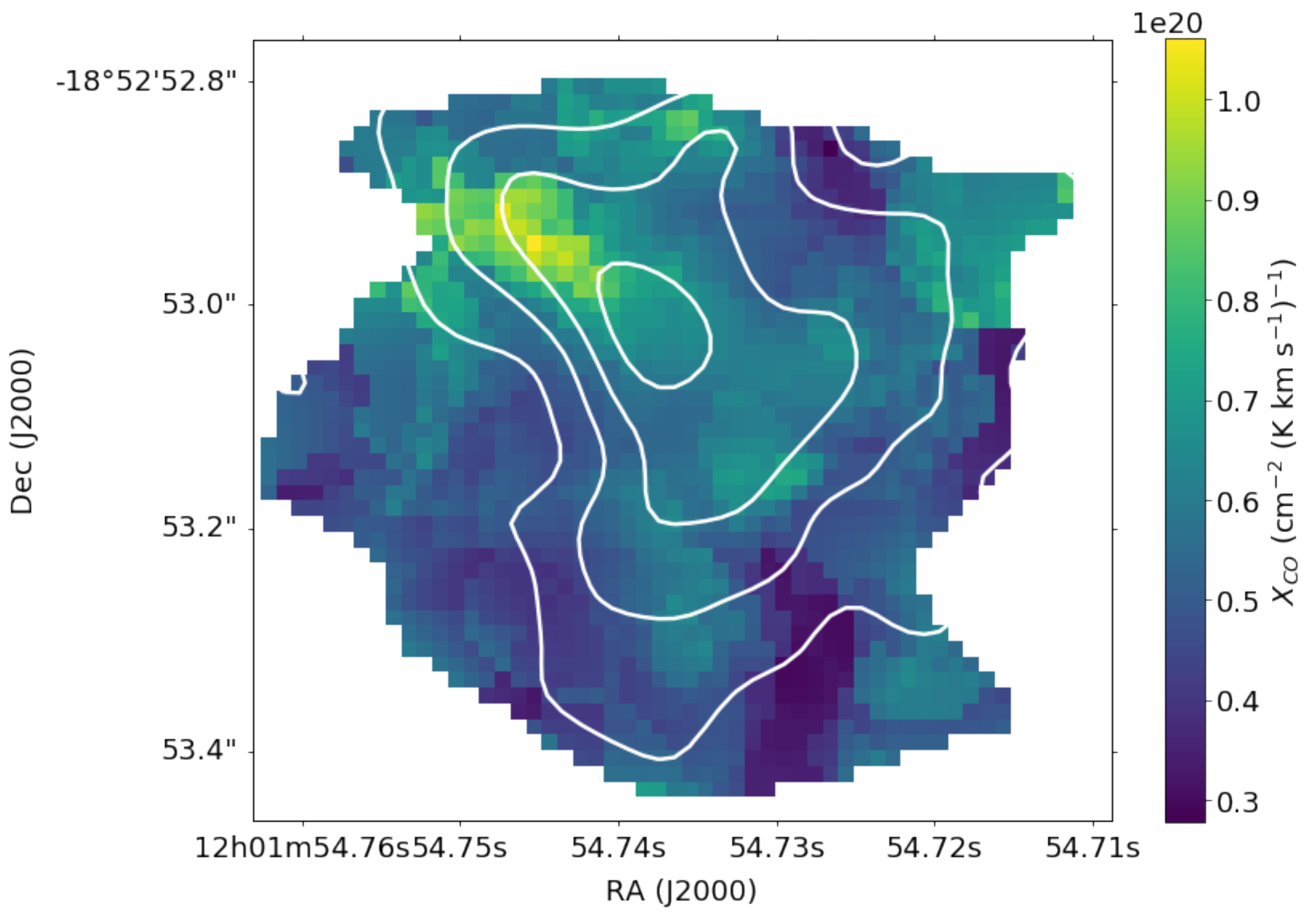}
    \includegraphics[width=0.38\textwidth]{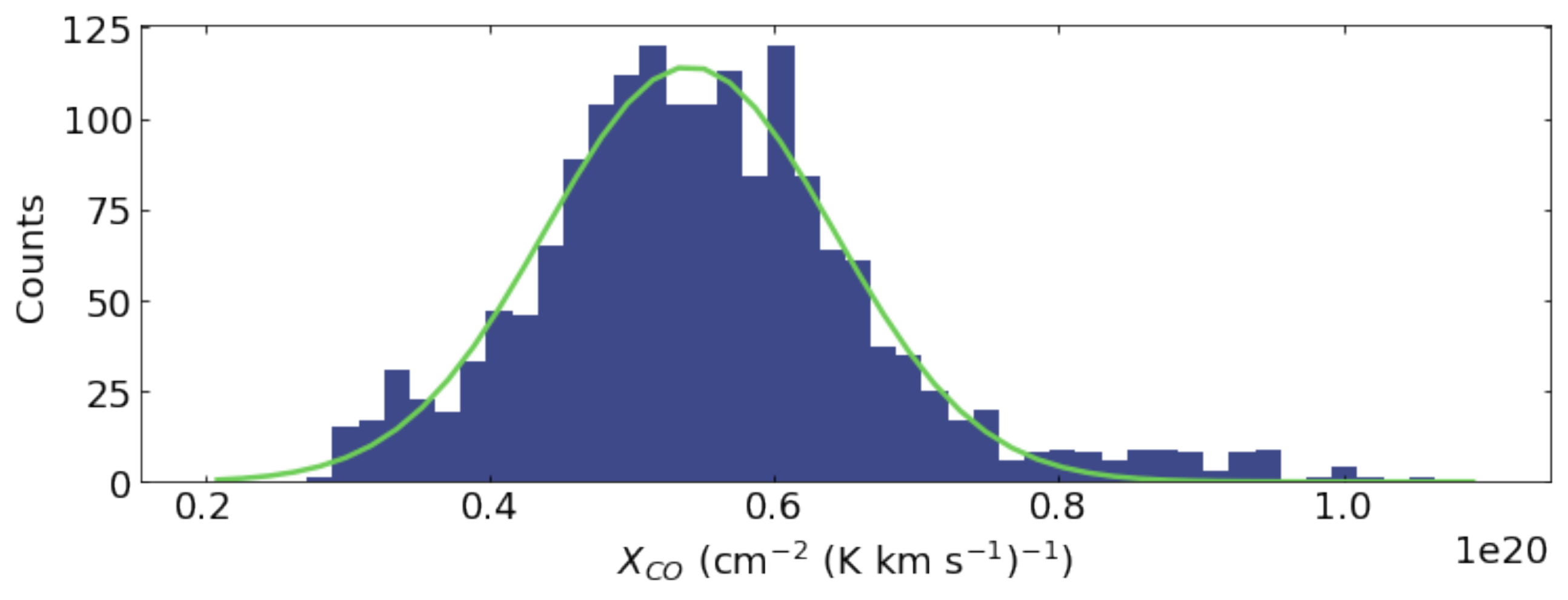}
    \caption{\emph{Top:} Spatial variation of the $X_{CO}$ factor with contours of \thirtCO(2-1) moment~0 overplotted. The values here are based on mass calculation assumptions of $X_{12} / X_{13}=70$ and H$_2$/\twelveCO$=10^{4.5}$, resulting in a total mass of $4.5\times10^6M_\odot$. Under these assumptions, $X_{CO}$ varies in the range $(0.3-1.1)\times10^{20}$ cm$^{-2}$ (K km s$^{-1}$)$^{-1}$ within this region. The values in this map will scale with different $X_{12}/X_{13}$ and H$_2$/\twelveCO\ assumptions, which results in a full range of $X_{CO} = (0.08-2.0)\times10^{20}$ cm$^{-2}$. \emph{Bottom:} Histogram of values in the plot above, with the fitted Gaussian overplotted. The distribution shows that there is a component of Gaussian noise with a mean value of  $X_{CO} = 0.54\times10^{20}$ cm$^{-2}$ and a width of  $X_{CO} = 0.1\times10^{20}$ cm$^{-2}$. However, the values on the upper and lower ends are higher than expected for only Gaussian noise, suggesting that these variations in measured $X_{CO}$ are due to physical variations, not just error in the measurements.}
    \label{fig:xcomap}
\end{figure}

We also see that this conversion factor appears to vary spatially over the Firecracker region. Figure\,\ref{fig:xcomap} shows a map of the derived $X_{CO}$ factor within the Firecracker region based on a map of the column density, and the map of the integrated line intensity of \twelveCO(2-1), which we assume is thermalized with respect to \twelveCO(1-0) (\twelveCO(2-1)/\twelveCO(1-0) = 1), and convert to K km s$^{-1}$ (Figure\,\ref{fig:moments}). Within such a map for a single set of $X_{12} / X_{13}$ and H$_2$/\twelveCO\ assumptions, the value of $X_{CO}$ varies by up to $\sim$80\% of the average across the Firecracker region. 

Furthermore, a histogram of the values for each map show that the distribution has a component of Gaussian noise, but the distribution at the higher and lower ends cannot be entirely explained by Gaussian noise in the measurements. These values are likely tracing physical variations in the conversion factor. When taking into account these spatial variations, as well as the range of mass estimates, we find that the conversion factor in the Firecracker can take on values in the range $X_{CO} = (0.08-2.0)\times10^{20}$ cm$^{-2}$.

\subsection{Column Density Radial Profile \label{subsec: NH2 profile}}

We examine the radial profile of the column density derived in Section\,\ref{subsec:mass} as a component of the mass estimate by calculating the azimuthal average of 1-pixel-wide (0.014") annuli around the center of the cloud as determined by the peak $N_{H_2}$ estimate. These annuli extend from a radius of 0.042" (4.5 pc) to the outer radius as determined by the 5$\sigma$ contour of the \thirtCO(2-1) (0.24", 26 pc). We measure this radial profile for column density estimates that assumed $X_{12} / X_{13}$ = 40, 70, 120, and 200. Since we only consider the column density normalized to the central peak when fitting the profile, assumptions of H$_2$/\twelveCO\ do not affect the fit.

To determine the physical nature of this internal structure, we compare it to the density profile of an isothermal, self-gravitating, pressure-confined sphere as described by \cite{Bonnor56} and \cite{Ebert55} and referred to as a Bonnor-Ebert profile. Starting with equations for hydrostatic equilibrium, an isothermal equation of state, and Poisson's equation, they arrive at a form of the Lane-Emden equation \citep{Chandra67}:

\begin{equation}
\frac{1}{\xi^2}\frac{d}{d\xi}\left(\xi^2\frac{d\psi}{d\xi}\right) = \exp{(-\psi)}
\end{equation}

\noindent In this equation, they have defined $\xi \equiv \left(\frac{4\pi G \rho_C}{a^2}\right)^{1/2} r$ as the dimensionless radial parameter, where $\rho_C$ is the central density and $a = \sqrt{\frac{k_B T}{\mu m_p}}$ is the isothermal sound speed. We have defined $\psi(\xi) \equiv \frac{\Phi_g}{a^2} = -\ln(\rho/\rho_C)$ as the dimensionless gravitational potential.

The Bonnor-Ebert profile is derived by numerically integrating this equation with the boundary conditions $\psi(0) = 0$ and $\frac{d\psi(0)}{d\xi} = 0$ to obtain a relation between $\rho/\rho_C$ and $\xi$. This density ratio can then be converted to a column density ratio with the assumption that the cloud is spherical. This profile is then fit to the observed structure profile by determining the best-fit values of $\xi_{max}$, the value of $\xi$ at the outer radius of the cloud. The resulting best fit for the derived set of column densities is characterized by $\xi_{max} = 3.4\pm0.4$. This fit has a $\chi^2$ value of 7.15, although when only the profiles with $X_{12} / X_{13}$ = 70, 120, and 200 are used, this fit changes to $\xi_{max} = 3.2\pm0.2$ with a $\chi^2$ value of 1.66. 

\begin{figure}
    \centering
    \includegraphics[width=0.45\textwidth]{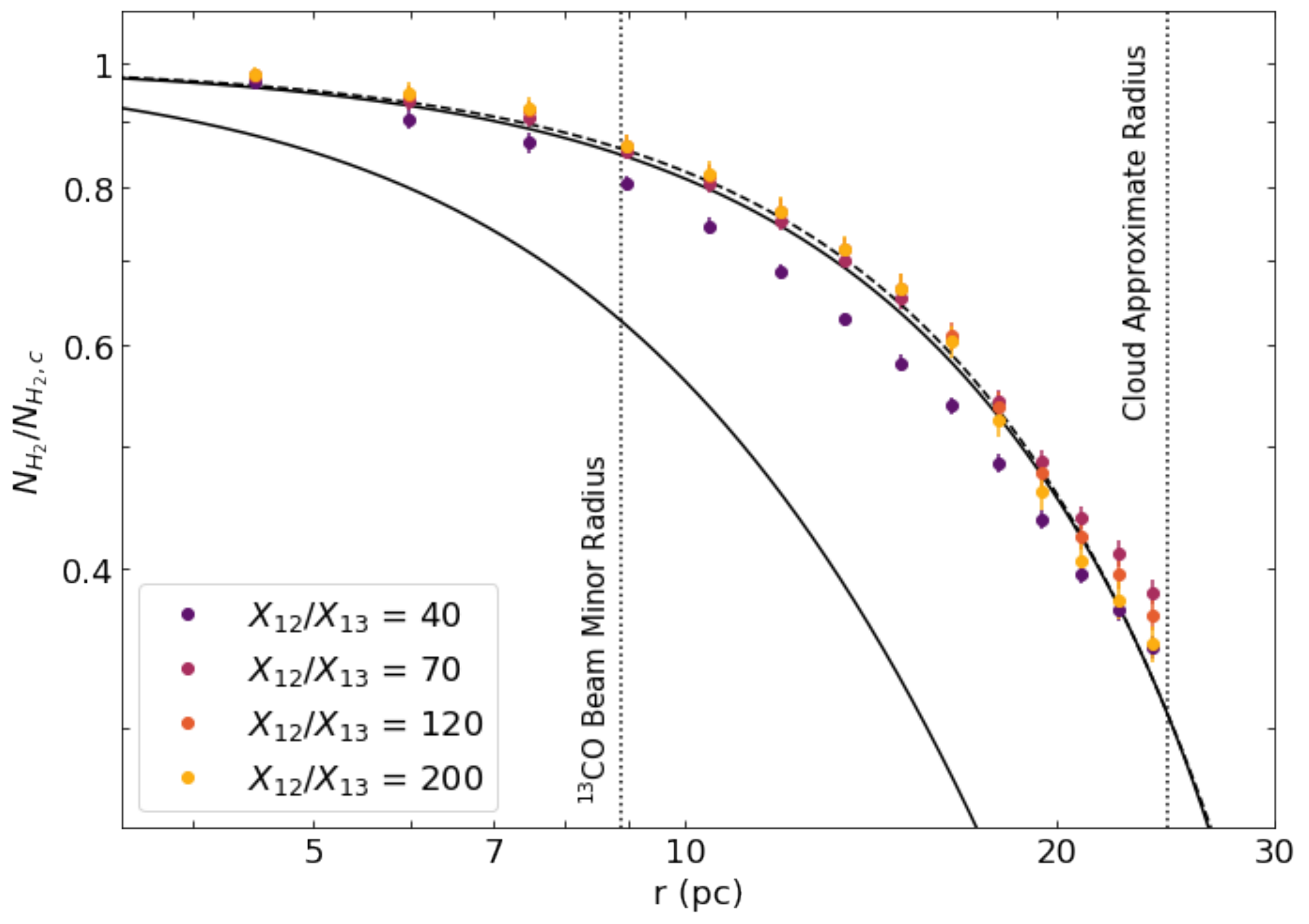}
    \caption{Radial profiles of the column density, $N_{H_2}$, normalized to the central column density, $N_{H_2,c}$, for different assumptions of $X_{12} / X_{13}$. Overplotted is the Bonnor-Ebert profile describing an isothermal, self-gravitating, pressure-confined sphere, with a fit characterized by $\xi_{max}=3.4\pm0.4$ (upper solid line), as well as the profile of the fitted gaussian (dashed line). The lower solid line is the Bonnor-Ebert profile with  $\xi_{max}=6.5$, indicating the profile below which the cloud would be gravitationally unstable. Since our best fit falls above this profile, the cloud is consistent with being gravitationally stable. 
     The error in the column density is taken to be the standard deviation in the azimuthal averaging. The dotted lines represent the radius of the \thirtCO(2-1) synthesized beam and the radius of the cloud. This Bonnor-Ebert fit has a $\chi^2$ value of 7.15.}
    \label{fig:BEprof}
\end{figure}

We note that in this fit, the cloud is resolved, but the points used in the fit are separated by less than the beam size, and so are correlated with each other. We also fit a circular Gaussian to the column density map, and plot the profile of this Gaussian in Figure\,\ref{fig:BEprof} as well. The cloud is consistent with both the Bonnor-Ebert profile and this Gaussian profile.

The  $X_{12} / X_{13}$ = 40 profile may appear separate from the other three profiles due to the slightly different morphology the mass map takes on when the \twelveCO\ is less optically thick, as is the case with this assumed lower abundance ratio. The other three profiles agree very well with each other. At the edge of the cloud, each profile begins to hit a noise threshold and so would be expected not to drop off as quickly as predicted, which agrees well with the observed profile.

The Bonnor-Ebert fit of the cloud's structure implies that it may be well-characterized as an isothermal, self-gravitating, pressure-confined sphere. We also note, however, that simulations of evolving star forming cores by \cite{Ballesteros2003} suggest that a Bonnor-Ebert profile can be mimicked by clouds that are not in hydrostatic equilibrium. This could be the case here for the Firecracker cloud, and so we are cautious in drawing conclusions about the cloud's physical state from this profile fit. The implications of this fit and concerns associated with it are addressed further in Section\,\ref{subsec: discuss P structure}.

\subsection{Cloud Pressure \label{subsec:pressure}}

We examine the effect of the cloud's environment on the parameters derived thus far by comparing the surface density, $\Sigma$, to a size-linewidth coefficient, $\sigma_V^2/R$. The velocity dispersions were determined by fitting two Gaussian profiles to the \twelveCO(3-2) emission line, one of which accounts for the second velocity component along the line of sight. The radius is taken to be the size of the aperture being measured, and the surface density is taken as the average within that aperture, based on the range of mass maps derived in Section\,\ref{subsec:mass}. 

\begin{figure*}
    \centering
    \includegraphics[width=0.34\textwidth]{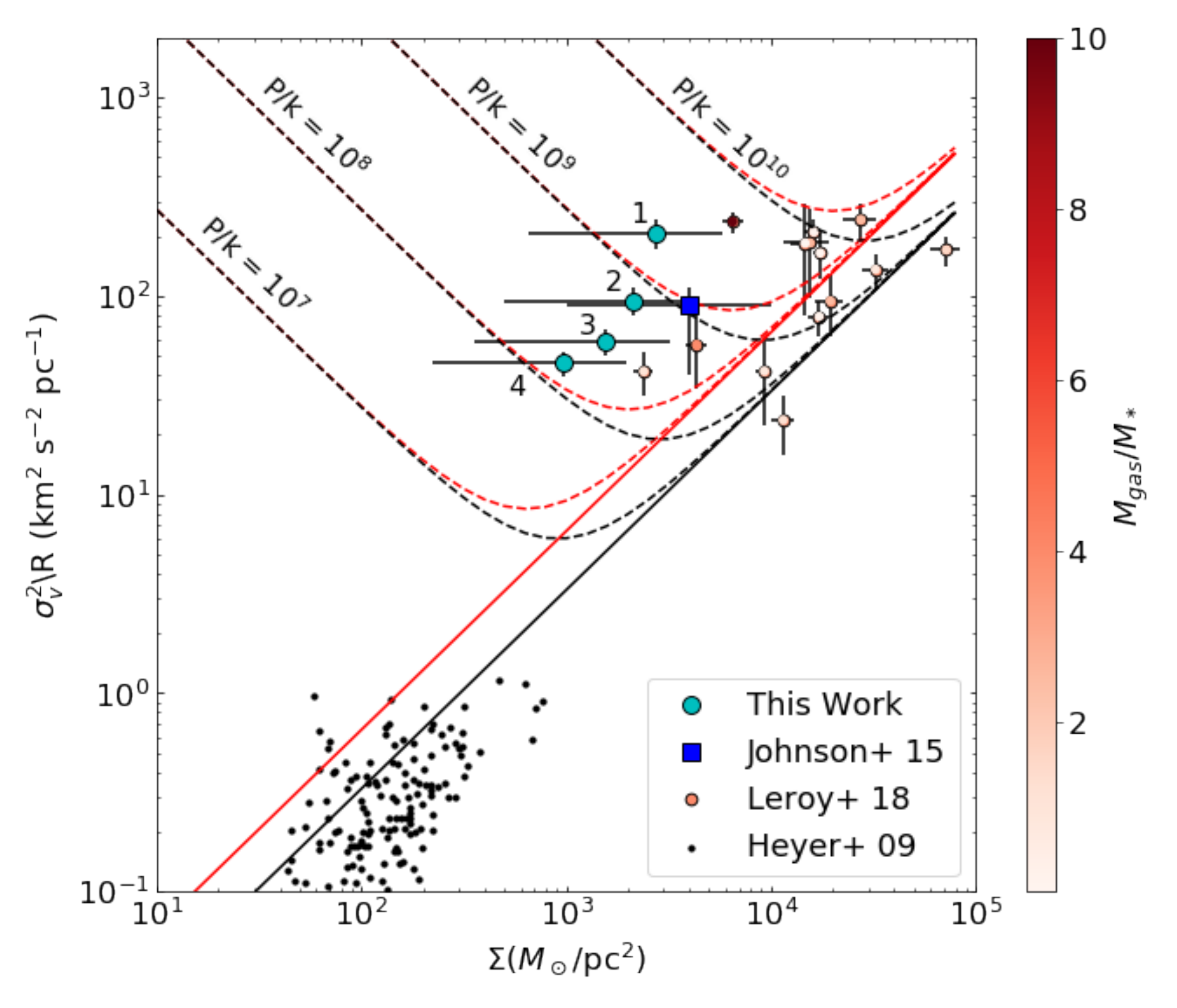}
    \includegraphics[width=0.33\textwidth]{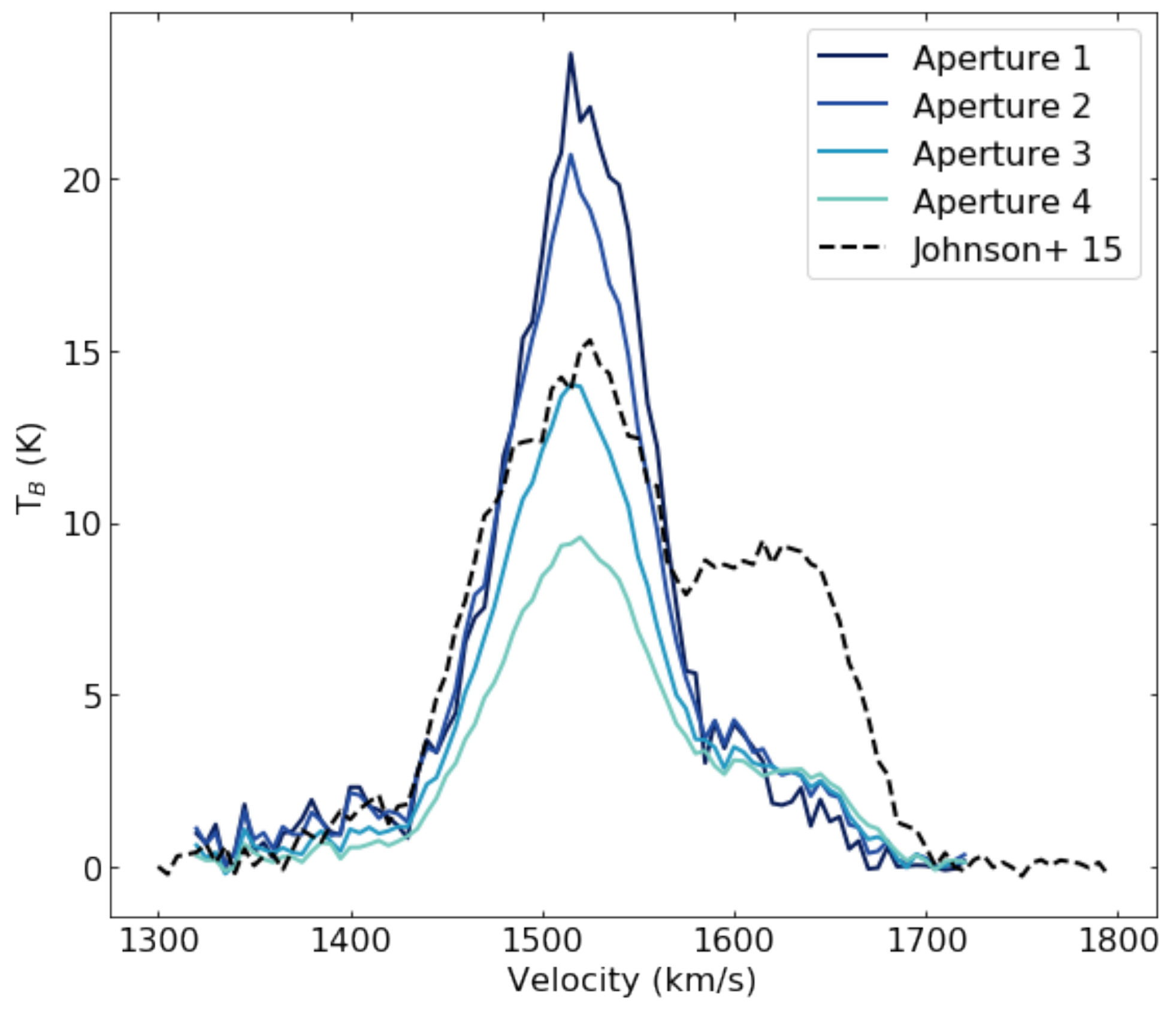}
    \includegraphics[width=0.28\textwidth]{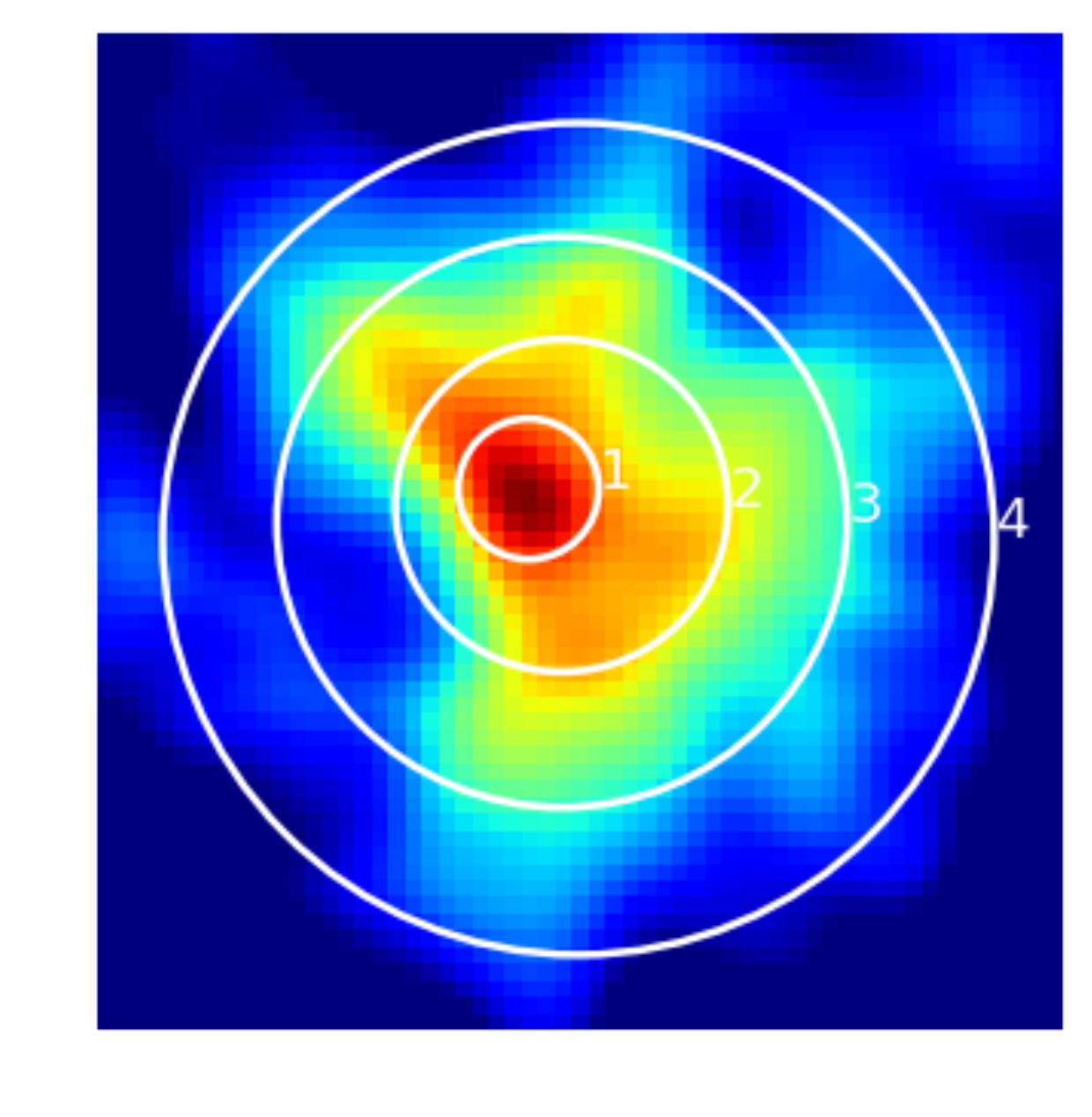}
    \caption{\emph{Left}: The size-linewidth coefficient ($\sigma_V^2/R$) and surface density ($\Sigma$) for the Firecracker cloud as measured in different apertures (cyan circles) and in previous analysis by \cite{J15} (blue square). Also shown are young massive clusters from \cite{Leroy18} in NGC 253, colored based on the ratio of $M_\text{gas}/M_*$, and typical molecular clouds observed in the Milky Way by \cite{Heyer09} for comparison (black circles). The black line corresponds to virial equilibrium, while the red lines correspond to free fall conditions \citep{Field11}. The position of the Firecracker cloud suggests that it is neither in virial equilibrium nor free fall, and so must be subject to a high external pressure ($\gtrsim10^8$ K cm$^{-3}$, dotted lines) to remain bound. This external pressure varies with the aperture used, which may indicate that we are seeing an internal pressure structure. Clusters with detected star formation from \cite{Leroy18} mostly fall along the virial or free fall lines, though the cluster with the highest $M_\text{gas}/M_*$ would require a high external pressure to be bound. This may suggest that the external pressure correlates with the evolutionary stage of the cluster. \emph{Center}: The \twelveCO(3-2) line profiles in each of the four regions as well as from previous analysis by \cite{J15} (dashed line). The linewidth remains approximately the same in each region, despite the large changes in peak brightness temperature. The larger contribution from the second velocity component in the \cite{J15} line profile is likely due to their larger synthesized beam (0.56"$\times$0.43" as compared to 0.16"$\times$0.15"). \emph{Right}: The four chosen apertures plotted on the \thirtCO(2-1) moment~0 map. These have radii of 0.06", 0.14", 0.24", and 0.35" for Apertures 1, 2, 3, and 4, which correspond to sizes of 6.4, 15, 26 and 37 pc respectively. }
    \label{fig:pressureplot}
\end{figure*}

We determine these parameters for four different apertures, shown in the right panel of Figure\,\ref{fig:pressureplot}. The largest, Aperture 4, is selected to include all \thirtCO(2-1) emission above $\approx4\sigma$, with a radius of 0.35" (37 pc). The next, Aperture 3, is selected to approximately match the size of the 5$\sigma$ contour of the \thirtCO(2-1), with a radius of 0.24" (26 pc). The next, Aperture 2, is selected to approximately follow the contour of 6$\sigma$ emission, with a radius of 0.14" (15 pc). Aperture 1 is approximately the \twelveCO(2-1) beam size, with a radius of 0.06" (6.4 pc). 

From the left panel of Figure\,\ref{fig:pressureplot}, these parameters indicate that the cloud is neither in virial equilibrium nor in free fall, implying that to be bound (as circumstantially suggested by its morphology), the cloud must be subject to a high external pressure with P$_e/k \gtrsim 10^8$ K cm$^{-3}$. This would agree with previous analysis by \cite{J15}, the fit of the Bonnor-Ebert profile in Section\,\ref{subsec: NH2 profile}, and theoretical expectations for cluster formation \citep{ElmegreenEfrefmov97}. 

Furthermore, we see that the inferred pressure increases as the aperture radius decreases, zooming in on the central region of the cloud. This may be an indication that we are tracing an internal pressure structure. It also may, however, be a measurement effect, since the radius of the selected aperture may not be a good indicator of the bound radius in the given region.

Also compared in the leftmost panel of Figure\,\ref{fig:pressureplot} young massive clusters discovered by \cite{Leroy18} in NGC 253, for which star formation has been detected. These clusters include both a gas and stellar mass component (with gas masses in the range $10^{3.6}-10^{5.7} M_\odot$, and stellar masses in the range $10^{4.1}-10^{6.0} M_\odot$), and we also compare the ratio $M_\text{gas}/M_*$ to their position in this plot. Most of the clusters fall along either the free fall or virial equilibrium lines, but one notable cluster with a significantly higher $M_\text{gas}/M_*$ than all of the other clusters is above these lines, suggesting a high external pressure (P$_e/k \gtrsim 10^9$ K cm$^{-3}$) would be required to keep it bound. Another cluster with a more modestly enhanced $M_\text{gas}/M_*$ is also above virial and free fall lines. This may be an indication that the pressure environment of massive clusters is correlated with the evolutionary stage of the cluster. This would support a scenario in which clusters form in high pressure environments, then the pressure dissipates or is dispelled as stars form and the cluster emerges.

Assuming the cloud is bound, we can also directly calculate what the expected external pressure would be from the cloud's mass $M$, its radius $R$, and its velocity dispersion $\sigma_V$ with the equation from \cite{Elmegreen89}:

\begin{equation}
    P_e = \frac{3\Pi M \sigma_V^2}{4\pi R^3}
\end{equation}

\noindent where $n_e = \Pi \langle n_e\rangle$, and we take $\Pi = 0.5$ \citep{J15}. If we calculate this pressure for Aperture 2 and a mid-range mass estimate, the mass within the aperture is $M = 1.5\times10^6 M_\odot$, $R = 15$ pc, and $\sigma_V = 38$ km s$^{-1}$, so the external pressure for this aperture would be P$_e/k = 4\times10^8$ K cm$^{-3}$. This measurement varies greatly with aperture selection and the mass estimate, and the full range of possible values is given in Table\,\ref{tab:cloud_params_derived}. This pressure range agrees with the values expected from Figure\,\ref{fig:pressureplot}.

\begin{table}
    \begin{center}
    \caption{Molecular Cloud Derived Properties}
    \begin{tabular}{c c c c c}
        \hline 
        \hline 
        Radius & M & T$_{Kin}$ & $n_{H_2}$ & P/$k$ \\
        (pc) & (10$^6 M_\odot$) & (K) & (cm$^{-3}$) & (K cm$^{-3}$) \\
        \hline 
         22 & 1--9 & 25--40 & 360-3150 & 0.5--22\x10$^8$ \\ 
        \hline
    \end{tabular}
    \end{center}
    \textbf{Notes.} The characteristic radius is the radius of a circle with the same area as that enclosed by the 5$\sigma$ contour of the \thirtCO(2-1) moment~0 map, with the assumption that the distance to
    the Antennae system is 22 Mpc. We also assume that $T_x \simeq T_{Kin}$.
    \label{tab:cloud_params_derived}
\end{table}

\subsection{Kinematics of the Local Environment}\label{subsec:PV}

To examine the larger local environment that may be causing the high external pressure derived above, we look at the kinematics of the surrounding region. If the source of pressure is ram pressure from the collision of molecular clouds, we might expect to see a ``broad bridge'' feature connecting the two clouds in the position-velocity diagram, as described by \cite{Haworth2015}. 

\begin{figure*}
    \centering
    \includegraphics[width=0.495\textwidth]{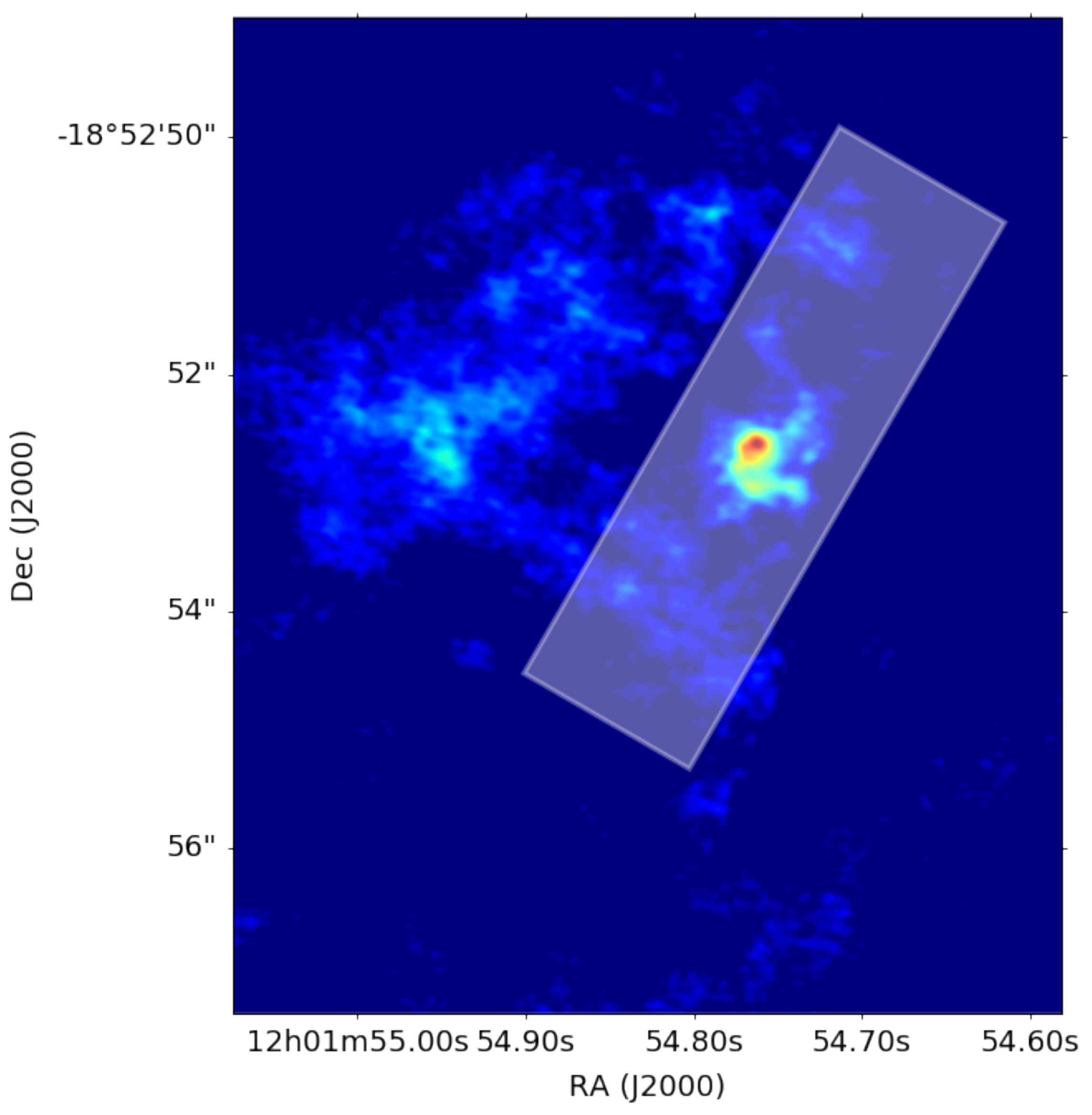}
    \includegraphics[width=0.49\textwidth]{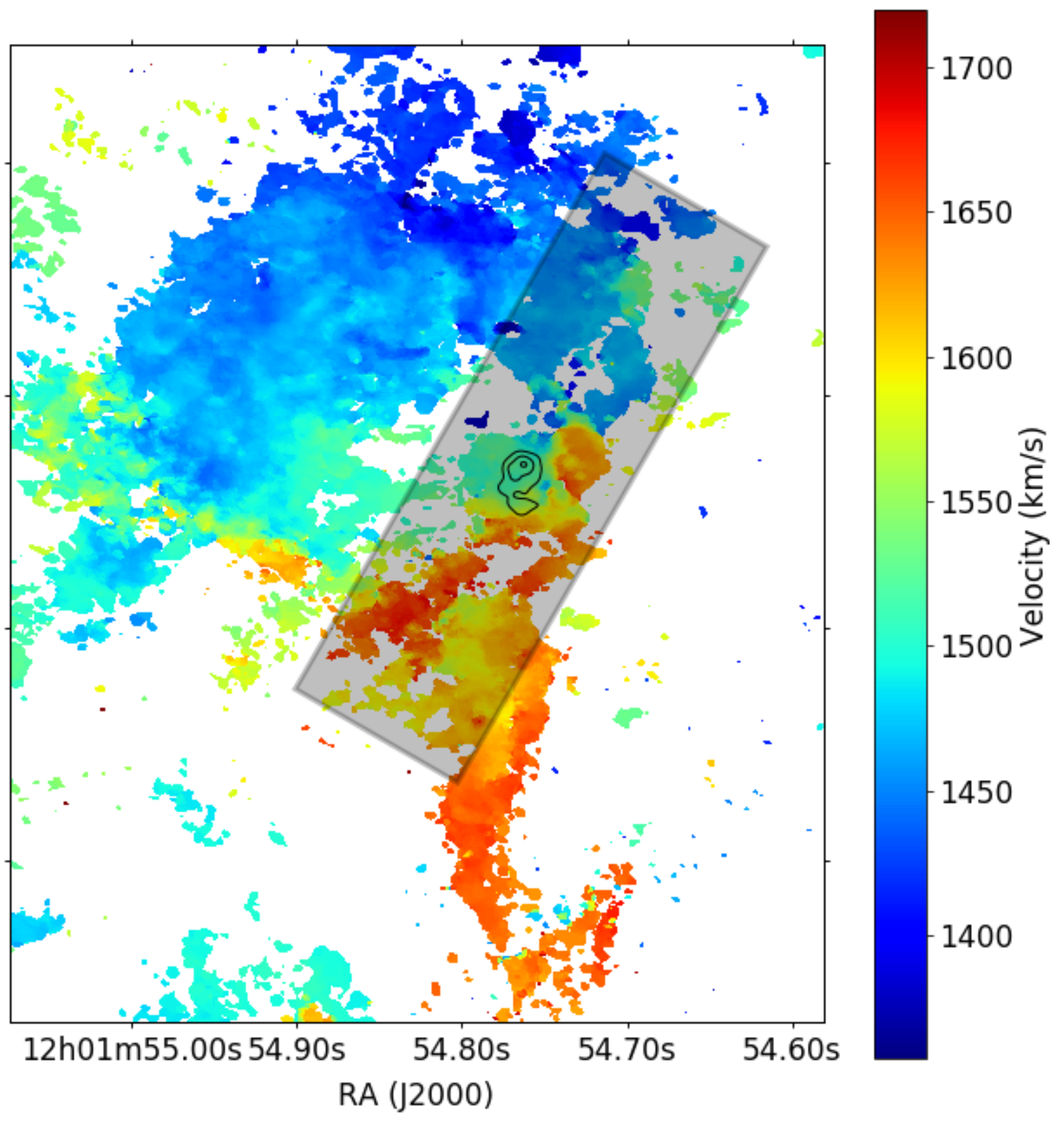}
    \includegraphics[width=\textwidth]{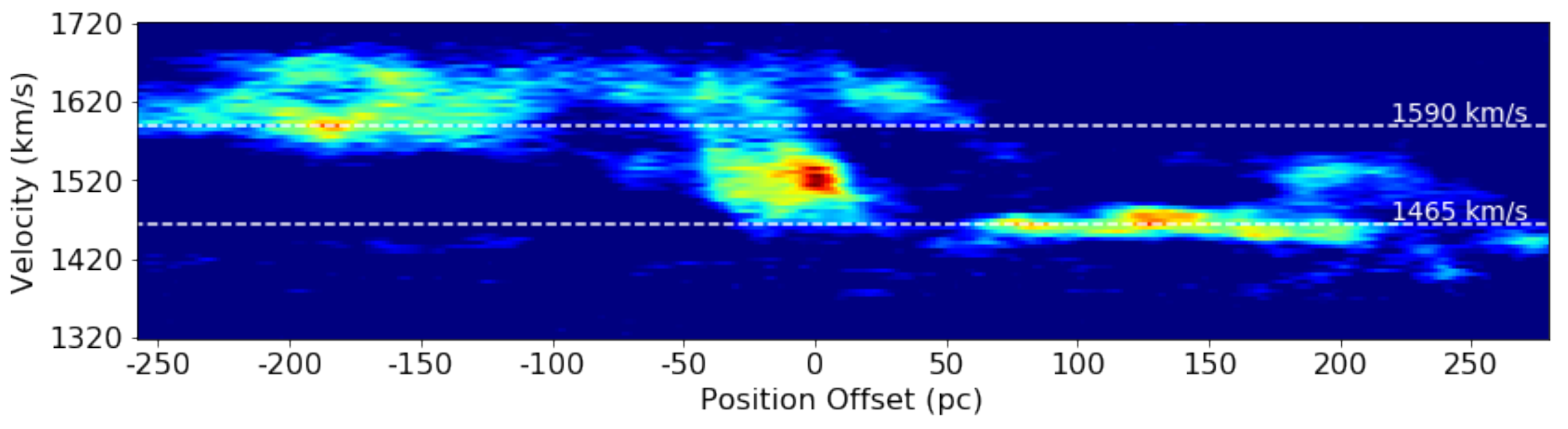}
    \caption{\emph{Top Left:} Moment~0 map of the Firecracker and the surrounding region, with the cut for the position-velocity diagram over plotted as a white rectangle. \emph{Top Right:} Moment~1 map of the same region, with the same cut shown as a gray rectangle. The cut was taken to roughly align with the greatest velocity gradient, which we expect would be the collision axis. Contours of the moment~0 of the Firecracker cloud are shown in black. \emph{Bottom:} Position-velocity diagram of the Firecracker from the cut shown above. The Firecracker cloud is seen in the center, and appears to be a `bridge' between the clouds on either side, from the top left of the plot to the bottom right. This bridge feature may be indicative of cloud-cloud collision, providing the high external pressure inferred in the Firecracker. The velocities of the colliding clouds are shown with the dotted lines, suggesting a relative collision velocity of $\sim 125$ km s$^{-1}$.}
    \label{fig:pvdiagram}
\end{figure*}

Using \twelveCO(2-1) emission of the Firecracker and the surrounding giant molecular cloud, we created a total intensity (moment~0) map by integrating over the velocity range 1430--1555 km s$^{-1}$, and a mean velocity (moment~1) map using a 0.6 mJy threshold. These maps were used to choose an angle and cut for a position-velocity diagram that would capture the proposed collision axis where the velocity gradient is greatest. These cuts and the resulting position-velocity diagram are shown in Figure\,\ref{fig:pvdiagram}.

The Firecracker does appear as a bridge between the two adjacent clouds, although it appears somewhat spatially separated from each. Its morphology is different from that of the broad bridge feature from \cite{Haworth2015}, but this may be due to a difference in viewing angle, as the line-of-sight of the simulated position-velocity diagrams from \cite{Haworth2015} were made directly along the collision axis. This would seem to suggest that cloud-cloud collision is likely occurring, and may be the source of pressure that we observe.

\subsection{HCN and HCO$^+$}\label{subsec:hcnhco}

HCN and HCO$^+$ are both tracers of dense gas. HCN has an optically thin critical density of $n_{crit} = 1.7\times10^5$ cm$^{-3}$ at 50 K, and an upper state energy of $T = 4.3$ K \citep{Shirley15}. HCO$^+$ has $n_{crit} = 2.9\times10^4$ cm$^{-3}$ at 50 K for optically thin gas and a very similar upper state energy of $T = 4.3$ K. Despite this similarity, these molecules do not appear to always be spatially correlated \citep{Johnson18}. This trend is continued in the Firecracker region. 

\begin{figure}
    \centering
    \includegraphics[width=0.45\textwidth]{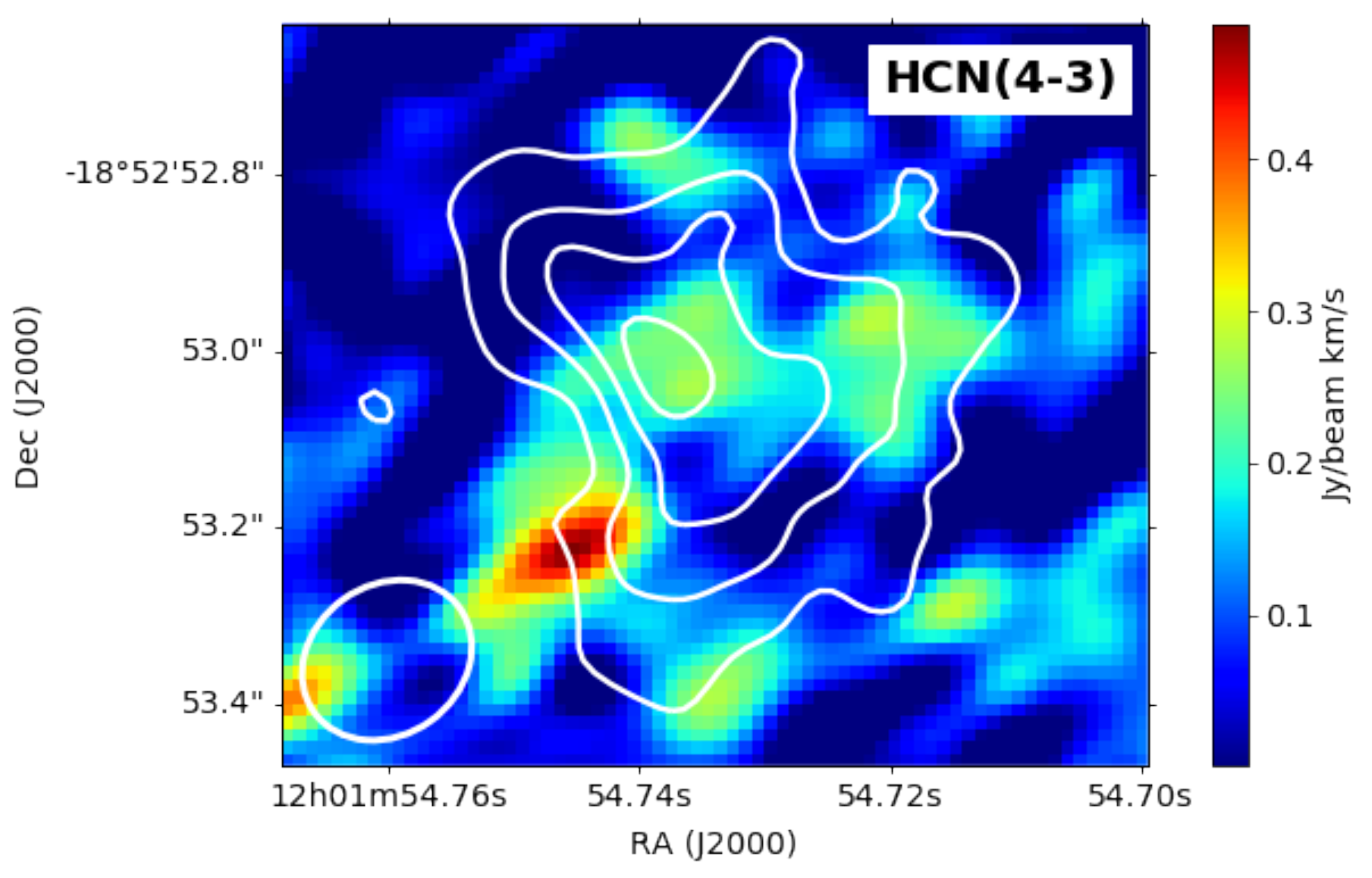}
    \includegraphics[width=0.45\textwidth]{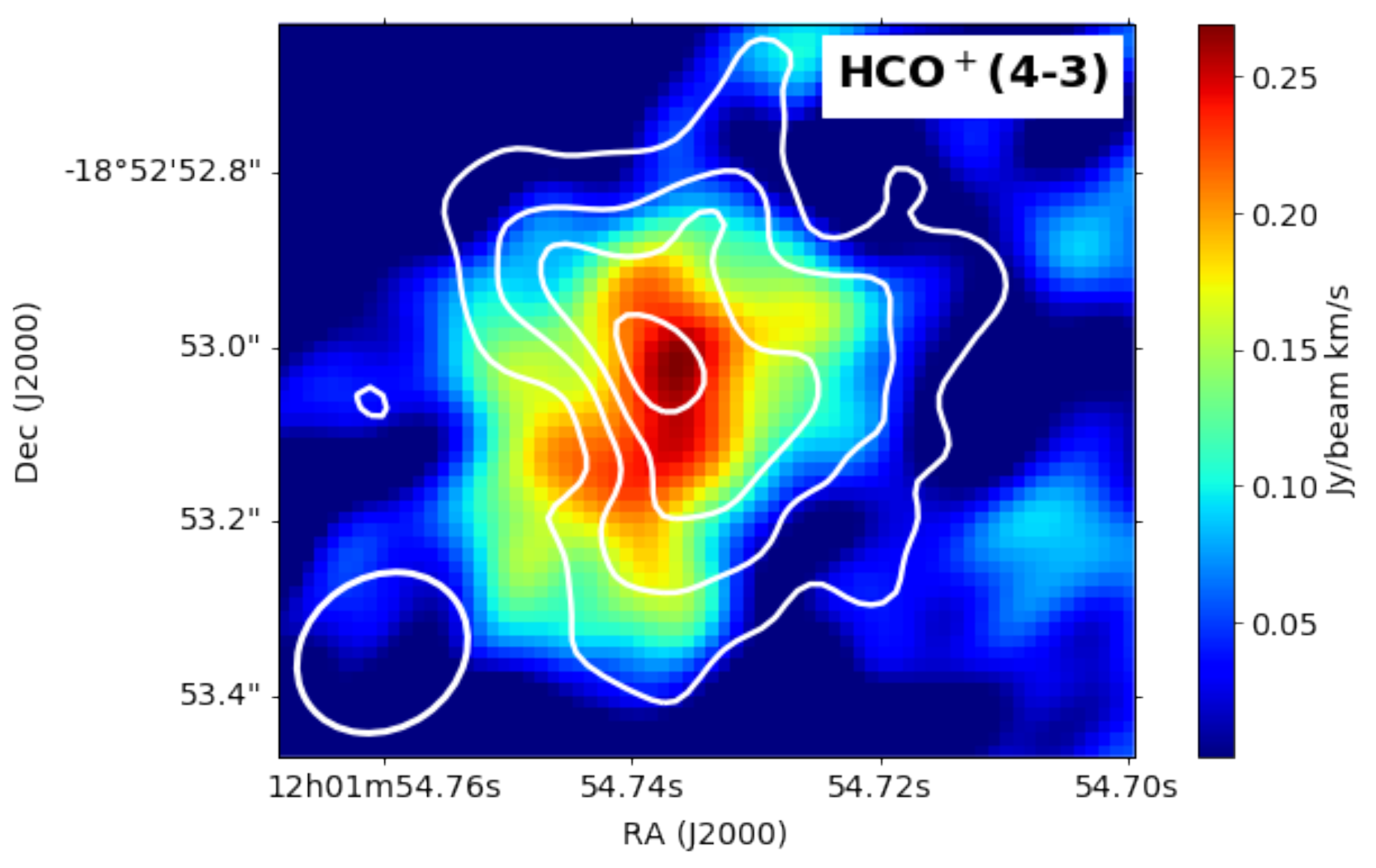}
    \caption{\emph{Top}: Total intensity (moment~0) map of HCN(4-3). \emph{Bottom}: Total intensity (moment~0) map of HCO$^+$(4-3). These total intensity maps were created by integrating across the velocity range 1430--1565 km s$^{-1}$. Both have contours of \thirtCO(2-1) moment~0 overplotted. Synthesized beams are 0.17"$\times$0.20" and 0.17"$\times$0.21" for HCN and HCO$^+$ respectively.}
    \label{fig:hcn/hco}
\end{figure}

Both of these species are weakly detected, with HCN(4-3) at the 4.1$\sigma$ level and HCO$^+$(4-3) at the 5.5$\sigma$ level in the total intensity (moment~0) map. The parameters of the data cube are given in Table \ref{tab:datacubes}. We created total intensity maps for each transition by integrating over the velocity range 1430--1565 km s$^{-1}$, and these are shown in Figure \ref{fig:hcn/hco}. In these images, it is apparent that the morphologies of the emission from these two molecules are quite different from each other (Figure\,\ref{fig:hcn/hco}).

Integrated fluxes from HCN(4-3) and HCO$^+$(4-3), as well as the \twelveCO(2-1), \twelveCO(3-2), and \thirtCO(2-1) lines, for each aperture described in Figure\,\ref{fig:pressureplot} are given in Table\,\ref{tab:fluxes}.

\begin{table}[]
    \begin{center}
    \caption{Integrated Fluxes of Emission Lines}
    \begin{tabular}{c|c c c c}
        \hline
         & Ap. 1 &  Ap. 2 &  Ap. 3 &  Ap. 4 \\
         \hline
         \hline
          S$_{^{12}CO(2-1)}$ & 0.96 & 4.5 & 9.3 & 14 \\
          S$_{^{12}CO(3-2)}$ & 1.7 & 8.3 & 18 & 26 \\
          S$_{^{13}CO(2-1)}$ & 0.04 & 0.19 & 0.50 & 0.90 \\
          S$_{HCN(4-3)}$ & 0.07 & 0.26 & 0.63 & 1.1 \\
          S$_{HCO^+(4-3)}$ & 0.07 & 0.30 & 0.64 & 0.78 \\
          \hline
    \end{tabular}
    \end{center}
    \textbf{Notes.} All integrated fluxes are measured in Jy km s$^{-1}$. Selected apertures are described in Section\,\ref{subsec:pressure}, shown in Figure\,\ref{fig:pressureplot}, and have radii of 0.06", 0.14", 0.24", and 0.35" for Apertures 1, 2, 3, and 4, which correspond to sizes of 6.4, 15, 26 and 37 pc respectively.
    \label{tab:fluxes}
\end{table}

From these moment~0 maps, we determine the average surface brightness of these two molecular lines in a 0.35" region around the Firecracker cloud (Aperture 4 in Figure\,\ref{fig:pressureplot}), is 33 K km s$^{-1}$ for HCN(4-3), and 22 K km s$^{-1}$ for HCO$^+$(4-3). These are much lower than the values measured in this region by \cite{Schirm16}, which are 46 K km s$^{-1}$ for HCN(1-0) and 73 K km s$^{-1}$ for HCO$^+$(1-0). These values from \cite{Schirm16} have been updated to account for a beam filling factor of 0.02 that we determine based on the cloud's now-resolved size of 0.21" and the \cite{Schirm16} beam of 1.52"$\times$1.86" for HCN(1-0) and 1.51"$\times$1.85" for HCO$^+$(1-0). This is expected, since \cite{Schirm16} present their measurements as upper limits due to the low resolution likely causing contamination from the surrounding region. We also expect that HCN(4-3) and HCO$^+$(4-3) are not thermalized with respect to HCN(1-0) and HCO$^+$(1-0), which further accounts for our lower surface brightness.

\begin{figure}
    \centering
    \includegraphics[width=0.45\textwidth]{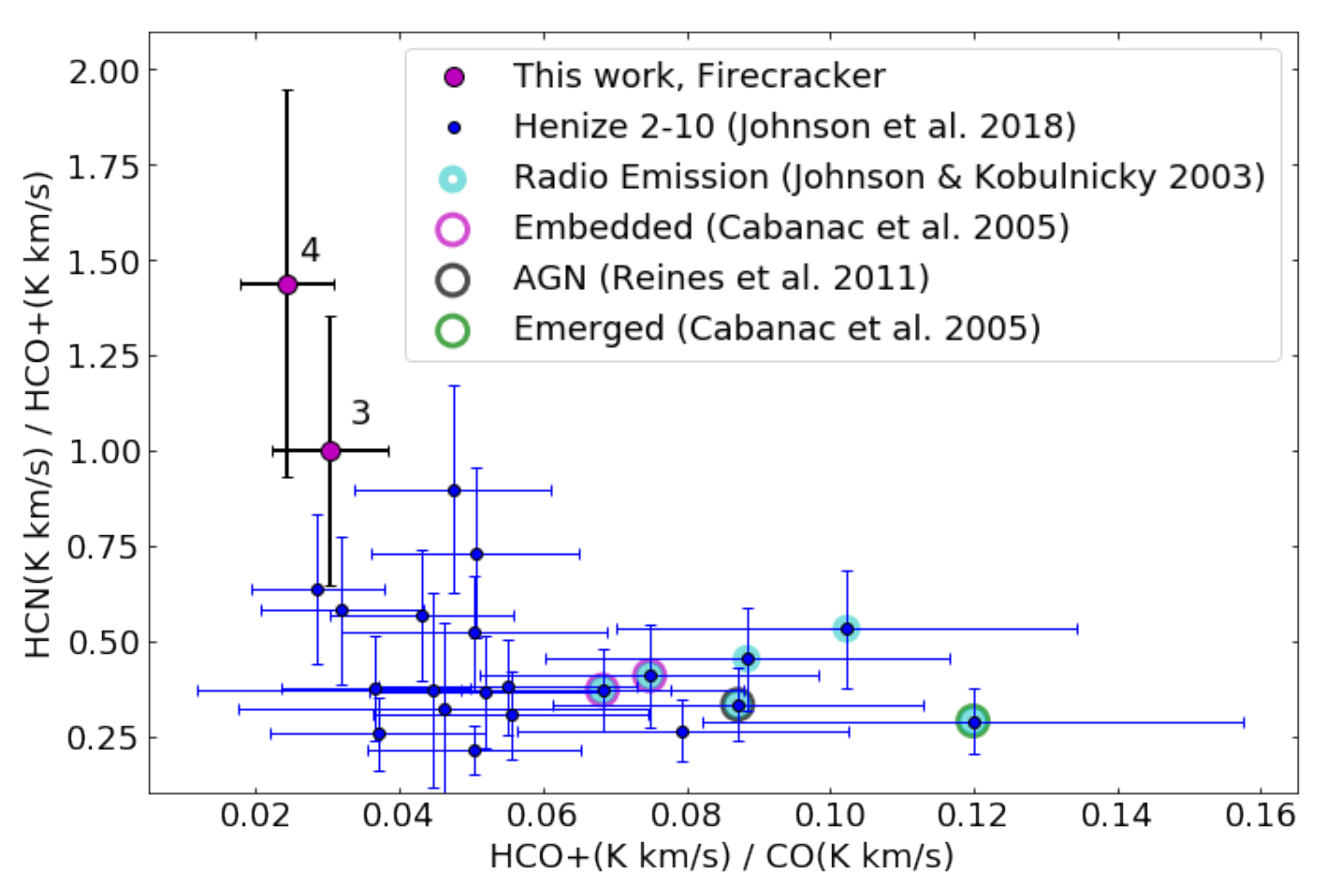}
    \caption{Ratios of HCN/HCO$^+$ and HCO$^+$/CO for natal SSCs in Henize2-10 from \cite{Johnson18} (blue) and the Firecracker (magenta). For the Firecracker, the HCO$^+$/CO ratio is a lower limit. Apertures for the Firecracker are defined as in Figure\,\ref{fig:pressureplot}, where Apertures 1 and 2 are not included since the synthesized beam is now larger than those regions. The trend in these regions from the pre-SSC Firecracker in the upper left to the older, star forming clusters in the bottom right suggests that HCN, HCO$^+$, and CO may be tracing cluster evolution.  }
    \label{fig:hcnevolution}
\end{figure}

We can next compare the relative line strengths in the Firecracker region to follow up on the analysis of \cite{Johnson18}, which suggested that HCN and HCO$^+$ strengths are associated with the evolution of proto-clusters. To do this, we convolve these images, as well as the \twelveCO(2-1) emission, to the same beam size (0.17"$\times$0.21") and look at the ratios of HCN(1-0)/HCO$^+$(4-3) and HCO$^+$(1-0)/\twelveCO(2-1), using the average surface brightness in K km s$^{-1}$ (Figure \,\ref{fig:hcnevolution}).
We look at these ratios in Apertures 3 and 4 as defined in the right panel of Figure\,\ref{fig:pressureplot}. Apertures 1 and 2 are not included since their radii are smaller than the synthesized beam for the new images.

We compare these ratios to those measured for HCN(1-0), HCO$^+$(1-0), and CO(2-1) in potential natal-SSCs found in the Henize 2-10 dwarf galaxy by \cite{Johnson18}. These natal-SSCs were selected via peaks in \twelveCO(2-1) emission, and 6 of the 21 regions had associated thermal radio emission \citep{JohnsonKobulnicky03}, indicative of stars having already formed. Two of these regions with associated radio emission are also shown to have high optical extinction, $A_V > 10$ \citep{Cabanac05}, implying that the clusters are still heavily embedded and have only recently formed stars. Another region exhibits nonthermal emission, suggesting a low-luminosity AGN \citep{Reines11} or an older evolutionary state, while another region includes several supernova remnants, has emerged from its surrounding gas, and has a cluster age of $>6$ Myr \citep{Cabanac05}. 

In Figure\,\ref{fig:hcnevolution}, we compare the two Firecracker apertures to those in Henize 2-10 and see that our new data supports the trend found by \cite{Johnson18}. As clusters evolve, these two line ratios appear to change, with regions that have already formed stars showing a higher ratio of HCO$^+$/CO and a lower HCN/HCO$^+$ ratio than regions at an earlier stage of evolution. 

Note that in this work we measure HCN(4-3) and HCO$^+$(4-3) rather than HCN(1-0) and HCO$^+$(1-0) as was used by \cite{Johnson18}. Due to the same upper state energies for HCN and HCO$^+$, we expect that the ratio of HCN/HCO$^+$ is not much affected by this difference, but the measured values of the ratio HCO$^+$/CO are lower limits for the Firecracker cloud.

We also note that the ratios seen in Figure\,\ref{fig:hcnevolution} differ for the two regions within the Firecracker, with values measured in Aperture 4 indicating a younger evolutionary stage than in Aperture 3. This seems to once again confirm the trend, as we would expect using a smaller aperture focused on the central region would include the gas most likely to begin star formation first, while the larger aperture includes more surrounding gas that has not yet begun evolving towards star formation. Thus the gas within Aperture 3 is expected to be at a more evolved state than the gas averaged within Aperture 4. 

For comparison, measurements in Henize 2-10 by \cite{Johnson18} used an aperture of 0.8", which at a distance of 9 Mpc corresponds to a physical size of 35 pc. This is approximately the same size as Aperture 4 in the Firecracker analysis.

\section{Discussion} \label{sec:discussion}

\subsection{Internal Structure}\label{subsec: discuss P structure}

In Figure\,\ref{fig:BEprof}, we show that the radial profile of the derived column density can be fit to a Bonnor-Ebert profile with $\xi_{max}=3.4\pm0.4$, implying that the cloud might be described by an isothermal, self-gravitating, pressure-confined sphere. 

If the fit is taken to be a true indication of the physics governing the cloud, then the value of $\xi_{max}$ that characterizes the best fit has further implications for the structure and state of the Firecracker cloud. It was shown by \cite{Bonnor56} that values of $\xi_{max} > 6.5$ are gravitationally unstable and will collapse. This would imply that our cloud, which is fit by $\xi_{max} = 3.4\pm0.4$, is still stable. This agrees with the results found in the left panel of Figure\,\ref{fig:pressureplot}, which show that the cloud is not experiencing free-fall collapse. This also further supports the belief that this cloud has not begun star formation and is an example of an undisturbed precursor cloud. 

Using the parameter $\xi_{max}=3.4$, we can also determine from the numerical solution of the Lane-Emden equation that the density contrast from the center to the boundary of the cloud would be $\rho_C/\rho_R = 3.2$. We can define a dimensionless mass, 

\begin{equation}
    m\equiv \frac{P_e^{1/2} G^{3/2} M}{a^4} = \left(4 \pi \frac{\rho_C}{\rho_R}\right)^{-1/2} \xi_{max}^2 \left.\frac{d\psi}{d\xi}\right\vert_{\xi_{max}}
\end{equation}

\noindent where $P_e$ is the bounding pressure, $M$ is the total mass, $a$ is the isothermal sound speed, and all of the values on the right side of the equation are known via the numeric solution to the Lane-Emden equation and the boundary condition of $\xi_{max} = 3.4$. We find that the dimensionless mass for the Firecracker cloud is $m=0.84$. This is less than the critical dimensionless mass derived by \cite{Bonnor56}, $m=1.18$, which is expected since we already demonstrated the cloud is stable. 

We can then derive from our definitions of $\xi$, $m$, and the isothermal pressure equation of state that

\begin{equation}
    a^2 = \frac{\xi_{max}}{m}\frac{GM}{R}\left(\frac{\rho_R}{\rho_C}\right)^{1/2}\left(\frac{1}{4\pi}\right)^{1/2}
\end{equation}

\noindent which allows us to determine $a$, the characteristic velocity of the equation of state in the Firecracker cloud. This velocity would be the isothermal sound speed, but in the Firecracker, we are likely dominated by microturbulence rather than thermal velocities. Depending on the value taken for the total mass, this velocity falls in the range $a = 10-30$ km s$^{-1}$. This is within a factor $\sim2$ to the velocity dispersion we measure in the cloud ($\sigma_V \sim 37$ km s$^{-1}$).

From this value for $a^2$, we can also derive another estimate of the external pressure confining the cloud, using an equation derived from the definition of $m$:

\begin{equation}
    P_e = \frac{m^2 a^8}{M^2 G^3}
\end{equation}

\noindent Taking the range of mass estimates, the external pressure in the Bonnor-Ebert profile would be $P_e/k = 0.05-4 \times 10^8$ K cm$^{-3}$, which agrees with the lower end of the pressure range determined in Section\,\ref{subsec:pressure}.

While these results do agree with expectations from the rest of our analysis, we also note that the profile is also consistent with a Gaussian profile, and also that it is possible for cores to be fit with a Bonnor-Ebert profile despite not actually obeying the physics of a stable, isothermal, pressure-confined sphere. \cite{Ballesteros2003} show that 65\% of the dynamic cores in their hydrodynamic models can be fit by Bonnor-Ebert profiles, and nearly half of these fits would suggest the dynamically evolving clouds are in hydrostatic equilibrium. Furthermore, their work shows that the parameters determined from the fits often varied from the actual values and depended on which projection of the core was used. We therefore are cautious in drawing firm conclusions from this fitted Bonnor-Ebert profile.

\subsection{High Pressure Environment}\label{subsec:discuss pressure}

To determine the source of the high external pressure implied by Figure\,\ref{fig:pressureplot} and the Bonnor-Ebert fit, we look to the encompassing cloud and its kinematics. \cite{J15} estimate that the weight of the surrounding super giant molecular cloud would only reach $P/k \sim 10^7$ K cm$^{-3}$. This falls short of the expected external pressure by one or two orders magnitude. 

One mechanism that may be able to increase the pressure in the region to the values we observe is ram pressure from colliding filaments. The Firecracker cloud is located at the confluence of two CO filaments identified by \cite{Whitmore14}, the region has a large velocity gradient across it, and is associated with strong H$_2$ emission \citep{Herrera11,Herrera12}. Work by \cite{Wei12} also shows that the overlap region may be dominated by compressive shocks. All of these would be consistent with collisions causing the high external pressure observed for the Firecracker cloud. Furthermore, the now-resolved irregular  structure of the cloud is consistent with the source of pressure being non-isotropic, as we would expect in the case of colliding gas filaments. This type of cloud-cloud collision has also been invoked as a trigger for massive star formation in several young clusters within the Milky Way and LMC, such as the Orion Nebula Cluster \citep[][and references therein]{Fukui2018}. \cite{Oey17} also see two kinematic components in a young SSC, which they suggest could be infall from cloud-cloud collision, or outflow due to feedback from the newly formed stars.

Examination of the position-velocity diagram (Figure\,\ref{fig:pvdiagram}) shows a hint of a ``broad bridge'' feature, described by \cite{Haworth2015} to be a signature of cloud-cloud collision. 
If this is indeed a case of cloud-cloud collision, we can use the $X_{CO}$ factor derived in Section\,\ref{subsec:Xco} to determine the density of the clouds on either side of the Firecracker. Taking the average value of $X_{CO} = 0.61\times10^{20}$ cm $^{-2}$ (K km s$^{-1}$)$^{-1}$, and assuming that the cross sections along the line of sight of the colliding clouds are twice the Firecracker's diameter, so have a depth of 88 pc, we find that the density of the colliding clouds is approximately $\rho \sim 10^{-21}$ g cm$^{-3}$ ($n_{H_2} \sim 220$ cm$^{-3}$).

From Figure\,\ref{fig:pvdiagram}, the velocities of the two colliding clouds are approximately 1465 km s$^{-1}$ and 1590 km s$^{-1}$, suggesting that the projected velocity difference at which they would be colliding is $v\sim 125$ km s$^{-1}$. Taking the ram pressure to be $P = \rho v^2$, this would imply that the pressure caused by such a cloud-cloud collision would be $P/k \sim 1.1\times10^9$ K cm$^{-3}$. While this is a fairly rough estimate of the ram pressure, it demonstrates that such a scenario would be capable of providing the high external pressures required for the Firecracker cloud to be bound.

Furthermore, this cloud-cloud collision scenario would imply that there will be a continued inflow of gas as the cloud begins to form a cluster of stars. Such an inflow would allow for accretion along filaments during the formation process, an important feature of the simulated cluster formation by \cite{Howard18}. This would support their suggestion that massive clusters can be formed by the same mechanisms that form smaller, less massive clusters.

\subsection{Comparisons to Other Molecular Clouds}\label{subsec: discuss comparisons}

To the best of our knowledge, the Firecracker is the only object that has been identified as having the properties necessary for SSC formation, while also having no detected thermal radio emission above a level of $N_\text{Lyc} \approx 6\times10^{50} \text{ s}^{-1}$, which corresponds to $\sim 60$ O-type stars, or $M_* \lesssim 10^4 M_\odot$. Given the cloud's mass of $M_\text{gas} = 1-9\times10^6 M_\odot$, this upper limit for stars formed is still at least two orders of magnitude less than the cloud's mass and so the Firecracker is likely to still be at a very early stage of cluster formation. Comparisons for this cloud must then come from SSCs that have detected star formation or galactic clouds that are forming less massive clusters that do not have the potential to form SSCs (where SSCs are expected to need $\gtrsim10^5 M_\odot$ to survive to be globular clusters).

\cite{Leroy18} identified a population of young massive clusters that have begun forming stars at detectable levels in NGC 253  ($N_\text{Lyc} > 5\times10^{50}$ s$^{-1}$), but most of which are still embedded in their natal material. They have gas masses in the range $10^{3.6}-10^{5.7} M_\odot$, stellar masses in the range $10^{4.1}-10^{6.0} M_\odot$, and FWHM sizes in the range 1.2-4.3 pc. Most of the clusters fall along the line for either virial equilibrium or free fall in the left panel of Figure\,\ref{fig:pressureplot}, but a couple with a notably higher $M_\text{gas}/M_*$ ratio are above these lines, suggesting that they would require a high external pressure to remain bound, similar to the Firecracker cloud. This may indicate a trend with evolution, since clusters at an early stage of formation will have turned less of their gas into stars. This would then support a scenario in which massive clusters are formed in high pressure environments, and then as stars form and the cluster evolves, the high pressure dissipates or is dispelled. 

We also note that the Firecracker shares several properties with the molecular cloud Sgr B2 in the CMZ of the galaxy. Sgr B2 has a mass of $8\times10^6 M_\odot$ and a diameter of 45 pc \citep{Schmiedeke2016}, which are well-matched to the Firecracker. It is also in a high pressure environment, with $P/k \sim 10^8$ K cm$^{-3}$ measured for embedded cores in the CMZ \citep{Walker2018}. High resolution ($\sim0.002$ pc) observations see this cloud break into several smaller clumps, the largest of which, Sgr B2 M, has a radius of $\sim 0.5$  pc, $M_\text{gas} \sim 10^4 M_\odot$, and $M_* \sim 1.5\times10^4 M_\odot$  \citep{Schmiedeke2016,Ginsburg2018}. This comparison suggests that at higher resolutions, we may see the Firecracker break into smaller protoclusters, which may or may not result in a single bound cluster. 

We also note however that the present day CMZ is a different environment from the Antennae overlap region, and the presence of a large SSC population in the Antennae \citep[$>$2,700 clusters with $M_*>10^5 M_\odot$;][]{Whitmore2010} and no SSCs in the Milky Way \citep[no young clusters with $M_*>10^5 M_\odot$;][]{PortZwart2010} also suggests that similar massive molecular clouds in the two regions could be expected to form different objects. 

\cite{Leroy18} also examine Sgr B2 as a comparison to their {massive clusters} and find that at their resolution of 1.9 pc, Sgr B2 would have a wider profile, narrower line width, and lower brightness temperature, suggesting it is a less dense version of the clouds forming clusters in NGC 253. These profiles extend to radii of $10$ pc, so cannot be directly compared to the Firecracker, since our resolution is $\gtrsim 10$ pc.

\subsection{Tracing Cluster Evolution}\label{subsec: discuss evolution}

\begin{figure}
     \centering
     \includegraphics[width=0.45\textwidth]{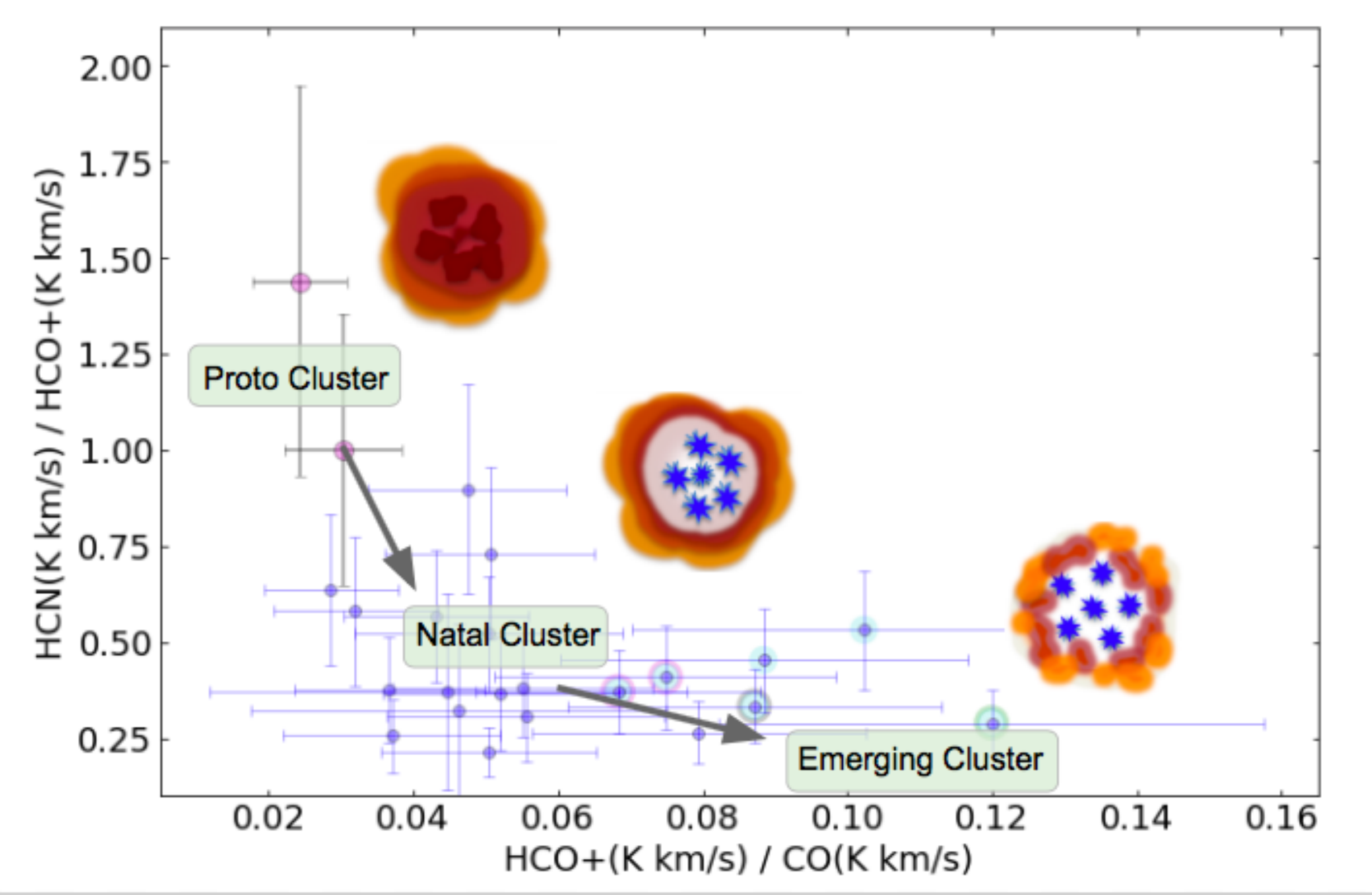}
     \caption{Ratios of HCN/HCO$^+$ and HCO$^+$/CO as shown in Figure\,\ref{fig:hcn/hco}, with a schematic showing the stages of evolution as suggested by the trend in the plot. HCN/HCO$^+$ appears to decrease with age, while HCO$^+$/CO appears to increase with age. This trend may allow us to use HCN and HCO$^+$ as a diagnostic of evolutionary state in unresolved cluster-forming systems.}
     \label{fig:evschematic}
 \end{figure}

 The trend shown in Figure\,\ref{fig:hcnevolution} indicates that both the ratios HCN/HCO$^+$ and HCO$^+$/CO are affected as the cluster evolves. HCN/HCO$^+$ appears to decrease with age, while HCO$^+$/CO appears to increase with age. This evolution is shown schematically in Figure\,\ref{fig:evschematic}.
 
 The mechanisms most likely to be driving the change in HCO$^+$/CO as discussed in \cite{Johnson18} are either the photo-enhancement of HCO$^+$ in the PDRs around newly formed stars \citep{Ginard12} or the dissociation of CO due to radiation from massive stars. The HCO$^+$ enhancement in these more evolved regions would also explain the decrease in HCN/HCO$^+$ that we observe. This trend could also be caused by a decrease in gas density as the star clusters evolve, causing the density of the gas to drop below the critical density of HCN while remaining higher than the critical density of HCO$^+$.
 
 If the increase in the HCO$^+$/CO ratio is due in part to the dissociation of CO as stars form, this would also be consistent with the analysis of \cite{Whitmore14}, which used CO brightness as a diagnostic of evolutionary stage for clusters. 
 
 \subsection{CO-to-H$_2$ Conversion Factor}\label{subsec: discuss XCO}
 
 In Figure\,\ref{fig:xcomap}, we see that $X_{CO}$ varies spatially by up to $\sim$80\% of the average within the Firecracker region, and that this variation is not solely a result of Gaussian noise in the measurements. $X_{CO}$ is expected to vary based on many parameters: high densities will cause $X_{CO}$ to increase, while high gas temperatures and super-virial velocity dispersions will lead to lower $X_{CO}$ values \citep{Bolatto13}. This results in a complicated picture for starbursting regions which experience all of these effects at once. Computational models by \cite{Narayanan11} demonstrate that $X_{CO}$ is lowest in regions of high SFR. 
 
 Comparing these predictions to the observed spatial variations in the Firecracker, we note that the central peak of the \thirtCO\ emission, where we expect initial collapse and star formation to occur, does not correspond to a particularly low or high value of $X_{CO}$ in the region as we might expect. Rather, the maximum and minimum locations occur around the edge of the cloud and do not appear to be correlated with the mass surface density, temperature, or velocity dispersion. This may simply be due to the complicated interplay of these three parameters' effects on $X_{CO}$.
 
 \subsection{Star Formation Efficiency}\label{subsec: discuss SFE}
 
 Another important parameter for studying the formation of globular clusters is the star formation efficiency (SFE), or how much of the gas in a molecular cloud is converted to stars. This value is defined as SFE = $M_{stars} / (M_{gas} + M_{stars})$. This parameter is important for the survival of the cluster, since much of the remaining gas will be dispersed after stars have formed. If the gas accounts for a large amount of the cluster's mass, the cluster will not remain bound after it has dispersed. Due to this type of argument, we have long believed that SFEs of $\sim$50\% are required to form globular clusters that last for $>10$ Gyr \citep{GeyerBurkert01}. However, more recent simulations have shown we may be able to relax that constraint to as little as $\sim$5\%, although a higher SFE makes a cluster more likely to remain bound \citep{Pelupessy12}. This value more closely matches values of $<10$\% for the SFE that has been measured in galactic clouds \citep{Evans09}.
 
 Recent computational models of massive star cluster formation show that there may be a correlation between the mass surface density, $\Sigma$, and the star formation efficiency. Both \cite{Kim18} and \cite{Grudic18} show clusters attaining SFEs of $\sim$50\% at high surface densities, $1300\ M_\odot$ pc$^{-2}$ in \cite{Kim18} and $3820\ M_\odot$ pc$^{-2}$ in \cite{Grudic18}.  These surface densities are well matched to those we measure in the Firecracker cloud, which range from $1000-6000 M_\odot$ pc$^{-2}$, depending on the assumed mass estimate. This has promising implications for the potential of the Firecracker to form a bound SSC. However, we also note that neither of these models take into account a high external pressure surrounding the precursor cloud as is inferred for the Firecracker for it to be bound. This external pressure would be likely to have strong implications for the outcome of the simulations, as the initial velocity dispersion of the gas would be much higher in this case.

Work done by \cite{Matthews18} attempts to observationally constrain the SFE, measuring the instantaneous mass ratio (IMR) in the Antennae overlap region. The IMR is an observational analog, defined as IMR = $M_{stars} / (M_{gas} + M_{stars})$, which would correspond to the SFE of an ideal system that had formed all its stars without yet dispelling its gas. \cite{Matthews18} find no correlation of IMR with surface density, and find that very few clusters in the region show an IMR greater than 20\% despite measuring surface densities up to $\sim10^4\ M_\odot$ pc$^{-2}$. This suggests the theoretical work may be optimistic in predicting SFEs in starbursting regions such as the Antennae galaxies.

\section{Conclusions}\label{sec:conclusions}

We present ALMA observations of the proto-SSC Firecracker cloud in the overlap region of the Antennae, looking at emission from \twelveCO(2-1), \twelveCO(3-2), \thirtCO(2-1), HCN(4-3), and HCO$^+$(4-3). These molecular lines were used to characterize the cloud and the surrounding environment at resolutions as low as $\sim$ 0.1" (10 pc). The findings are summarized below.

\begin{itemize}
    \item We determine the mass of the cloud to be in the range 1--9$\times 10^6 M_\odot$ and its characteristic radius is 22 pc. These both agree with previous measurements by \cite{J15} and are consistent with the cloud having the potential to form a super star cluster.
    
    \item We do not detect continuum emission at any of the three observed frequencies. This allows us to put an upper limit on the mass (9$\times10^6 M_\odot$), as well as constrain abundance ratios of \twelveCO/\thirtCO\ and H$_2$/\twelveCO\ within this region. Certain combinations of these two ratios are disallowed by this upper mass limit. 
    
    \item We calculate the CO-to-H$_2$ conversion factor and determine that it varies spatially by up to $\sim$80\% of the average, with average values in the range $X_{CO} = (0.12-1.1)\times10^{20}$ cm$^{-2}$ (K km s$^{-1}$)$^{-1}$. This is consistent with $X_{CO}$ values typically adopted in starburst regions. The spatial variations cannot be explained solely by Gaussian noise in the measurements, and do not align with areas expected to have the greatest SFR. Instead, the variations likely depend on a complex combination of temperature, density, and velocity dispersion.
    
    \item We find that the radial profile of the column density can be fit by a Bonnor-Ebert profile characterized by $\xi_{max} = 3.4\pm0.4$, and that this profile is also consistent with a Gaussian profile. The Bonnor-Ebert fit would suggest that the Firecracker cloud might be described as an isothermal, self-gravitating, pressure-confined sphere, similar to those forming clusters in our galaxy on smaller scales. This profile would also suggest that the cloud is gravitationally stable. We caution though that simulations of dynamic clouds not in equilibrium have also been shown to be fit by Bonnor-Ebert profiles, which may be the case here for the Firecracker cloud.
    
    \item We determine from surface density and size-linewidth parameters that the cloud is not in free-fall or virial equilibrium, and so must be subject to a high external pressure, $P/k \gtrsim 10^8$ K cm$^{-3}$, if it is a bound structure. A comparison with young massive clusters in NGC 253 that have detected star formation suggests a potential trend in which clusters with a low $M_\text{gas}/M_*$ ratio (and so are likely more evolved) are near virial equilibrium or free fall, while clusters with a higher $M_\text{gas}/M_*$ would require similar high pressures to remain bound. This would agree with theoretical predictions that high pressure environments are necessary for cluster formation. It also agrees with the Bonnor-Ebert fit's prediction that the cloud is pressure-bound and gravitationally stable.
    
    \item The position-velocity diagram of the Firecracker and its surrounding cloud shows what may be a ``broad bridge'' feature, which is indicative of cloud-cloud collision. An estimate of the density and relative velocity of the colliding filaments suggests that they are capable of producing a ram pressure of $\sim 1.1\times10^9$ K cm$^{-3}$, consistent with the high pressures needed for the cloud to be bound.
    
    \item We demonstrate that the Firecracker cloud further supports the findings of \cite{Johnson18} that HCN and HCO$^+$ appear to trace the evolutionary stage of clusters. As stars begin to form, the HCN/HCO$^+$ ratio decreases while the HCO$^+$/CO is enhanced. This could be due to some combination of enhancement of HCO$^+$ in PDRs as stars form, dissociation of CO from massive stars, and changes in gas density as the cluster evolves.
    
    \item The measured surface density range of $\Sigma = 1000-6000 M_\odot$ pc$^{-2}$ may indicate that the cloud is capable of a star formation efficiency as high as $\sim$50\%. A high SFE is predicted to be necessary for a globular cluster to form and remain bound throughout its lifetime.
    
\end{itemize}\

\acknowledgements

We thank the anonymous referee whose helpful comments improved this manuscript. This research is supported by NSF grants 1413231 and 1716335 (PI: K.~Johnson). This paper makes use of the following ALMA data: ADS/JAO.ALMA\#2015.1.00977.S and ADS/JAO.ALMA\#2016.1.00924.S. ALMA is a partnership of ESO (representing its member states), NSF (USA) and NINS (Japan), together with NRC (Canada), NSC and ASIAA (Taiwan), and KASI (Republic of Korea), in cooperation with the Republic of Chile. The Joint ALMA Observatory is operated by ESO, AUI/NRAO and NAOJ. The National Radio Astronomy Observatory is a facility of the National Science Foundation operated under cooperative agreement by Associated Universities, Inc.. The research of WEH and CDW is supported by grants from the Natural Sciences and Engineering Research Council of Canada. CDW also acknowledges support from the Canada Research Chairs program and the Canada Council for the Arts.

\bibliographystyle{yahapj}
\bibliography{references.bib}

\end{document}